\pdfoutput=1
\documentclass{nature}
\usepackage[backend=bibtex,style=nature,defernumbers=true]{biblatex}
\usepackage{lineno}
\usepackage[T1]{fontenc}
\usepackage[final]{microtype}
\usepackage{lmodern}
\usepackage{graphicx}
\usepackage{caption}
\usepackage{multirow}
\usepackage{aas_macros}
\usepackage[english]{babel}
\usepackage{csquotes}
\usepackage{longtable}
\usepackage{amssymb}
\usepackage{grffile}

\DeclareCaptionLabelSeparator{bar}{ | }
\defbibheading{bibliography}{%
}
\patchcmd{\thebibliography}{\section*{\refname}}{}{}{}

 \makeatletter
 \newcommand*{\newbibstartnumber}[1]{%
   \apptocmd{\thebibliography}{%
     \global\c@NAT@ctr #1\relax
     \addtocounter{NAT@ctr}{-1}%
   }{}{}%
 }
 \makeatother

 \makeatletter
 \patchcmd{\blx@citation@entry}
   {\ifinlistcs{#1}{blx@segm@\the\c@refsection @\the\c@refsegment}
      {}
      {\listcsgadd{blx@segm@\the\c@refsection @\the\c@refsegment}{#1}}}
   {\ifboolexpr{ test {\blx@ifentryseen@global{#1}}
      or test {\ifinlistcs{#1}{blx@segm@\the\c@refsection @\the\c@refsegment}} }
      {}
      {\listcsgadd{blx@segm@\the\c@refsection @\the\c@refsegment}{#1}}}
   {}{}
 \makeatother

\def\llm{\textsc{LLmodels}}

\def\logg{\log(g)}
\def\teff{T_{\rm eff}}

\def\kms{km~s$^{-1}$}
\def\bs{\langle B \rangle}

\def\synth3{\textsc{Synth3}}

\def\vsini{\upsilon\sin i}

\def\b{|\mathrm{\mathbf{B}}|}

\def\btimesfi{\sum|\mathbf{B}|_if_i}

\def\Msun{M_{\odot}}

\newcommand{\abn}[1]{\alpha(\mathrm{#1})}

\ifdefined\ion
\renewcommand{\ion}[2]{\textup{#1\,\textsc{\uppercase{#2}}}}
\else
\newcommand{\ion}[2]{\textup{#1\,\textsc{\uppercase{#2}}}}
\fi

\def\lamlam{\lambda\lambda}
\newcommand*\degr{\ensuremath{^\circ}}

\def\stokesi{Stokes-\ensuremath{\rm I}}
\def\stokesv{Stokes-\ensuremath{\rm V}}

\title{Strong dipole magnetic fields in fast rotating fully convective stars}

\author{D.~Shulyak$^{1*}$, A.~Reiners$^1$, A.~Engeln$^1$, L.~Malo$^{2,3}$, R.~Yadav$^4$, J.~Morin$^5$, \& O.~Kochukhov$^6$}

\begin{document}

\maketitle

\begin{affiliations}
 \item[$^1$] Institute for Astrophysics, Georg-August University, Friedrich-Hund-Platz 1, D-37077 G\"ottingen, Germany
 \item[$^2$] Canada-France-Hawaii Telescope (CFHT) Corporation, 65-1238 Mamalahoa Highway, Kamuela, Hawaii 96743, USA
 \item[$^3$] Institut de recherche sur les exoplan\'etes (iREx), D\'epartement de Physique, Universit\'e de Montr\'eal, Montr\'eal, QC H3C 3J7
 \item[$^4$] Harvard-Smithsonian Center for Astrophysics, 60 Garden Street, Cambridge, MA 02138, USA
 \item[$^5$] LUPM, Universite de Montpellier, CNRS, place E. Bataillon, F-34095 Montpellier, France
 \item[$^6$] Department Physics and Astronomy, Uppsala University, Box 516, 751 20, Uppsala, Sweden
\end{affiliations}

\begin{refsegment}
\defbibfilter{notmain}{not segment=\therefsegment}

\begin{abstract}
M dwarfs are the most numerous stars in our Galaxy with masses between approximately $0.5$ and $0.1$ solar mass.
Many of them show surface activity qualitatively 
similar to our Sun and generate flares, high X-ray fluxes, 
and large-scale magnetic fields\supercite{1995ApJ...438..269B,1984ApJ...279..763N,2013SAAS...39.....C,2012LRSP....9....1R}. 
Such activity is driven by a dynamo powered by the convective motions 
in their interiors\supercite{1981ApJ...248..279P,1984ApJ...279..763N,2003A&A...397..147P,2011ApJ...743...48W,2014ApJ...794..144R}. 
Understanding properties of stellar magnetic fields in these stars finds a broad application
in astrophysics, including, e.g., theory of stellar dynamos and environment conditions 
around planets that may be orbiting these stars.
Most stars with convective envelopes follow a rotation-activity relationship where various 
activity indicators saturate in stars with rotation periods 
shorter than a few days\supercite{1984ApJ...279..763N,2003A&A...397..147P,2014ApJ...794..144R}.
The activity gradually declines with rotation rate in stars rotating more slowly. 
It is thought that due to a tight empirical correlation between X-ray and magnetic flux\supercite{2003ApJ...598.1387P}, 
the stellar magnetic fields will also saturate, to values around $\sim4$~kG\supercite{2009ApJ...692..538R}.
Here we report the detection of magnetic fields above the presumed saturation limit in four fully convective M-dwarfs. 
By combining results from spectroscopic and polarimetric studies
we explain our findings in terms of bistable dynamo models\supercite{2013A&A...549L...5G,2015ApJ...807L...6G}: 
stars with the strongest magnetic fields are those in a dipole dynamo state, while stars in a multipole
state cannot generate fields stronger than about four kilogauss.
Our study provides observational evidence that dynamo in fully convective M dwarfs 
generates magnetic fields that can differ not only in the geometry of their large scale component,
but also in the total magnetic energy.
\end{abstract}

Our understanding of origin and evolution of the magnetic fields in 
M dwarfs is based on the models of the rotationally driven
convective dynamos. Modern observations provide two important constraints for these models. 

First, from the analysis of circular polarization in spectral lines
we infer that large-scale magnetic fields
tend to have simple axisymmetric geometry with dominant poloidal
component in stars that are fully convective. In contrast, M dwarfs that are hotter
and therefore only partly convective tend to have more complex fields with strong
toroidal components\supercite{2010MNRAS.407.2269M}. 
However, there is a number of exceptions when a rapidly-rotating fully convective star 
generates a large-scale magnetic field with a complex multipole geometry.
This dichotomy of magnetic properties in stars that have similar
stellar parameters may be explained in terms of dynamo bistability:
stars can relax to either dipole or multipole states
depending on the geometry and the amplitude 
of an initial seed magnetic field\supercite{2013A&A...549L...5G,2015ApJ...807L...6G}.
Note, however, that dynamo bistability was observed only in models of stars
with masses $M\leqslant0.2\Msun$.

The second observational constraint is the rotation--activity relation\supercite{1984ApJ...279..763N,2007AcA....57..149K,2014ApJ...794..144R,2017ApJ...834...85N}.
A remarkable feature of this relation is the
existence of two branches, a saturated and a non-saturated branch. 
In the non-saturated branch, the amount of non-thermal (e.g., X-ray) emission generated
by the star grows with rotation rate. On the saturated branch (corresponding rotation periods shorter than $\sim4$ days)
the level of activity shows only little dependence on rotation. 

In observations of solar active regions and some young stars, absolute
X-ray luminosity was found to be proportional to magnetic flux\supercite{2003ApJ...598.1387P}
($4\pi R^2 \bs$, i.e., surface area of the star times magnetic flux density).
This correlation suggests that, as long as X-ray luminosity saturates in fast rotating
stars, a similar saturation of the magnetic flux (and/or magnetic flux density) may also take place.
A growing database of stellar magnetic field measurements 
showed that the strength of the maximum possible surface magnetic field 
reaches values around $3-4$~kG 
in the coolest M dwarf stars\supercite{1985ApJ...299L..47S,1996ApJ...459L..95J,2007ApJ...656.1121R,
2009ApJ...692..538R,2010ApJ...710..924R}.
Mean fields stronger than this have not been detected in any low-mass star,
which was viewed as an evidence for the magnetic field saturation\supercite{2009ApJ...692..538R}.
It occurs for stars with rotational periods shorter than a few days
implying that the generation of magnetic flux itself does not grow beyond a
level proportional to the bolometric luminosity, as supported by theoretical studies\supercite{2009Natur.457..167C}. 
Strong kilogauss level magnetic fields are equally found in stars with dipole and multipole
states suggesting that these stars may share a common mechanism of saturation.

{Measurements of the magnetic fields in M dwarfs usually utilize either unpolarized light (\stokesi),
or circularly polarized light (\stokesv). The important difference between the two is that \stokesv\ carries information
about polarity and geometry of the magnetic field which is organized at large spatial scales on the stellar surface. 
However, \stokesv\ can not see small scale magnetic fields because these fields have different polarities
and thus their contribution is canceled out when the star is observed from a large distance as a point source.
{To the contrary}, both large and small scale fields contribute to the unpolarized \stokesi\ light.
Thus, using \stokesi\ one can measure the strength of the total magnetic field\supercite{2016ASSL..439..223M}.}
However, all previous studies made use primarily of either \stokesi\ or \stokesv\ radiation 
without detailed analysis of the link between the two. For instance, it was known\supercite{2009A&A...496..787R}
that \stokesv\ measurements can miss up to $90$\% of the total magnetic flux density of the star, while for many
stars with published \stokesv\ magnetic maps only coarse measurements of 
the magnetic field strength from \stokesi\ were available.

The analysis of the magnetic fields 
has significantly improved over the last decade thanks to development of new methods and techniques,
making it possible to look at magnetic properties of stars in more detail.
In our research we used data collected over several years with the twin spectropolarimeters ESPaDOnS and NARVAL
mounted at the $3.6$~m Canada-France-Hawaii Telescope (CFHT) and the $2$~m Telescope Bernard Lyot (TBL) 
at the Pic du Midi (France), respectively\supercite{2014PASP..126..469P}. 
Both instruments cover a spectral range between $367-1050$~nm and have spectral resolution $R\approx65,000$ (in polarimetric mode).
We chose a sample of M dwarfs
with known rotational periods and available magnetic maps derived from inversion of \stokesv\ 
spectra\supercite{2010MNRAS.407.2269M,2017ApJ...835L...4K}. 
The ultimate aim of our work is to measure magnetic flux density from 
independent spectroscopic diagnostics by utilizing up-to-date radiative transfer modelling.
Contrary to previous studies we analyze \stokesi\ 
spectra because we want to capture the total magnetic flux which provides 
essential constraints for the underlying dynamo mechanism.

Our analysis resulted in the detection of very strong magnetic fields in four M dwarf stars.
{We report the strongest average magnetic field $\bs\approx7.0$~kG in the star WX~UMa (Gl~412~B),
a field of $\bs\approx6.0$~kG in stars Wolf~47 (Gl~51) and UV~Cet (Gl~65~B), 
and a field about $5.0$~kG in V374~Peg, respectively.}
Note that WX~UMa is the only cool active star known to date where the Zeeman splitting in single atomic lines
is clearly observed at optical wavelengths (Zeeman splitting at long near-infrared wavelengths is easier 
to see and therefore it has already been detected in some objects\supercite{1985ApJ...299L..47S,2009AIPC.1094..124K}).
Figure~1 demonstrates example fits to magnetically sensitive spectral lines
in these and in three other magnetic M dwarfs that we have chosen for comparison.
For WX~UMa we measure a minimum magnetic field of about $6$~kG from Zeeman splitting 
in \ion{Rb}{i}~$\lambda794.76$~nm and \ion{Ti}{i}~$\lambda837.7$~nm lines, 
and larger values when applying radiative transfer model to fit full line profiles.

WX~UMa is an unusual object:
it is a rapid rotator (rotation period $P=0.74$d) but it has
relatively low inclination angle $i=40$\degr\ of its rotation axis 
with respect to the observer. This results in relatively small projected rotational velocity 
$\vsini\approx6$~\kms\ and corresponding {rotation} broadening of spectral lines become
weaker than the magnetic one.
This is why we could detect the Zeeman splitting in atomic lines in this star. 
{For instance, in Wolf~47 we also measure a very strong field, but high $\vsini\approx11$~\kms\ makes it impossible
to observe corresponding Zeeman splitting in this star (see Fig.~1). 
In our final measurements of magnetic fields we use numerous lines of FeH molecule at $\lamlam990-995$~nm (Wing-Ford $F^4\,\Delta-X^4\,\Delta$ transitions)
and near-infrared Ti lines located between $960-980$~nm. The details of our analysis are given in section Methods.
Thus, we report for the first time  a definite detection of the magnetic fields in M dwarfs well beyond its recent maximum value
of about four kilogauss. Our Fig.~2 demonstrates that new detection
make a clear difference because they substantially extend the range of measured to date 
magnetic fields in stars with saturated activity level.}

Our discovery allows us to draw several important conclusions.
First we note that, according to published magnetic field maps, WX~UMa and Wolf~47
belong to the dipole branch and show concentration of strong magnetic fields around rotation poles 
(so-called magnetic polar caps)\supercite{2010MNRAS.407.2269M}.
In contrast to this, none of the multipole stars with similar parameters and rotation rates
exhibits a field stronger than $\bs\sim4$~kG.
Thus our analysis implies that stars with dipole states may generate stronger fields 
compared to stars with multipole states at the same rotation rate. 
This conclusion is in agreement with the predictions 
of bistable dynamo models\supercite{2013A&A...549L...5G,2015ApJ...807L...6G}.

Next, we emphasize that the magnetic field we derive in stars with a dipole dominated geometry
does not necessarily reflect the surface-averaged magnetic flux density.
Indeed, because of geometric projection, we observe fields stronger than the surface average field
in stars oriented to us with their magnetic poles,
and weaker fields is stars that rotate equally fast but seen equator-on.
From the magnetic maps corresponding to dynamo models\supercite{2015ApJ...813L..31Y} 
we estimate that WX~UMa with its $7.3$~kG magnetic field and inclination of $i=40$\degr\ would 
show ``only'' $6.4$~kG should it had the inclination $i=60$\degr, which matches well enough
the field that we measure in, e.g., Wolf~47 which has $i=60$\degr\supercite{2010MNRAS.407.2269M}.
The magnetic flux density averaged over whole stellar surface for WX~UMa amounts now to $6.8$~kG
(against measured $7.3$~kG),
which is still significantly higher than the maximum field strength in stars with multipole regime.
(See our Supplementary Fig.~1 for the visualization of the geometrical effect).
Note that this geometrical effect should be absent in stars with multipole fields 
because these fields are randomly distributed over stellar surface without any preferred axis.

{By analyzing the dependence between measured magnetic field and rotation
period we observe that the magnetic flux density in stars with dipole-dominated geometry 
still grows with rotation rate.}
In Fig.~3 we plot our magnetic field measurements for all
M dwarfs with known magnetic geometries\supercite{2010MNRAS.407.2269M,2017ApJ...835L...4K}.
We find the same trend of increasing magnetic field strength as rotation periods decrease
to about four days (so-called unsaturated regime), as was found in previous studies\supercite{2009ApJ...692..538R}. 
However, we now show
that magnetic fields in M dwarfs with multipole dynamo state saturate below periods of about four days
with the saturated magnetic flux density $\bs\sim4$~kG
(corresponding Rossby number $Ro=P/\tau_{\rm c}<0.1$, where $P$ is the stellar rotation period and $\tau_{\rm c}$
is the convective turnover time, see Supplementary Fig.~2),
{while some stars with dipole state exhibit surface fields unambiguously above $4$~kG
and demonstrate no obvious saturation effect. However, from our limited data set 
it remains unclear whether these stars saturate at
faster rotation or if the saturation value for the magnetic field is only
stronger, in comparison to stars with multipole state.
Nevertheless, from this plot we can see that dynamo processes may behave differently with rotation rate in dipole 
and multipole branches.}
{It is thus important to address this question with future observations in more detail.}

An alternative explanation of the dichotomy in magnetic field geometries
is to assume that M dwarfs can have magnetic cycles
during which magnetic field can change its geometry and strength over cycle period.
This would explain the difference in magnetic fields
between individual objects.
Note that the signs of activity cycles were found in some M dwarfs stars
based on photometric/spectroscopic observations\supercite{2013ApJ...764....3R},
including a recent detection of a clear $7$~year photometric cycle in fully convective and slowly rotating ($P\approx83$d) 
M dwarf Proxima Centauri\supercite{2017MNRAS.464.3281W}. 
By computing a model for this star it was shown that dynamo in fully convective M dwarfs 
can indeed generate \textit{magnetic cycles} with repeated reversals of the magnetic polarities at different
latitudes with time, provided that these stars rotate slow enough to develop differential rotation in their convection zones\supercite{2016ApJ...833L..28Y}.
On the other hand, same models showed that as rotation increases, the generated magnetic fields finally become strong
enough to quench the differential rotation and thus no magnetic cycles are expected:
once reached, a star never changes its dynamo state.
This theoretical prediction is supported by ZDI observations of fast rotating ($Ro<0.1$) M dwarfs, though magnetic cycles
could be as long as decades in these stars\supercite{2015MNRAS.449.3471S}.
Thus, to test which of the two scenarios (i.e., bistable dynamo vs. cyclic dynamo) 
is responsible for the observed magnetic properties of M dwarfs a long-term spectropolarimetric monitoring is required.

\printbibliography[segment=\therefsegment,resetnumbers=false]
\end{refsegment}

\begin{addendum}
\item 
This project was carried out in the framework of the DFG funded
CRC~963~--~Astrophysical Flow Instabilities and Turbulence (projects
A16 and A17). We acknowledge support from the DAAD strategic partnership project U4-Network to D.S.
and funding through a Heisenberg
Professorship RE 1664/9-2 to A.R.
{R.K.Y is supported by NASA Chandra grant GO4-15011X.}
O.K. acknowledges funding from the Swedish Research Council and the Swedish National Space Board.
We also acknowledge the use of electronic
databases (VALD, SIMBAD, NASA's ADS). This research is based on
observations collected at the Canada-France-Hawaii Telescope (Hawaii) 
and Telescope Bernard Lyot (France).

\item[Author Contribution]
D.S. contributed general scientific ideas and conclusions, 
carried out the processing, modelling, and analysis of observed data.
A.R. contributed general scientific ideas and conclusions.
A.E. carried out modelling and analysis of observed data.
L.M. carried out telluric correction on selected targets.
R.Y. provided theoretical dynamo models and magnetic maps.
J.M. provided and analyzed additional Stokes-V data on some ESPaDOnS targets.
O.K. contributed to development of atomic line analysis methodology.
All authors contributed to the text of the paper and discussed the results.
\item[Author Information] 
Reprints and permissions information is available at
www.nature.com/reprints. The authors declare no competing financial
interests.  Readers are welcome to comment on the online version of
the paper. Correspondence and requests for materials should be
addressed to D.S. (denis@astro.physik.uni-goettingen.de).
\end{addendum}

\begin{figure}
  \centerline{
    \includegraphics[width=\hsize]{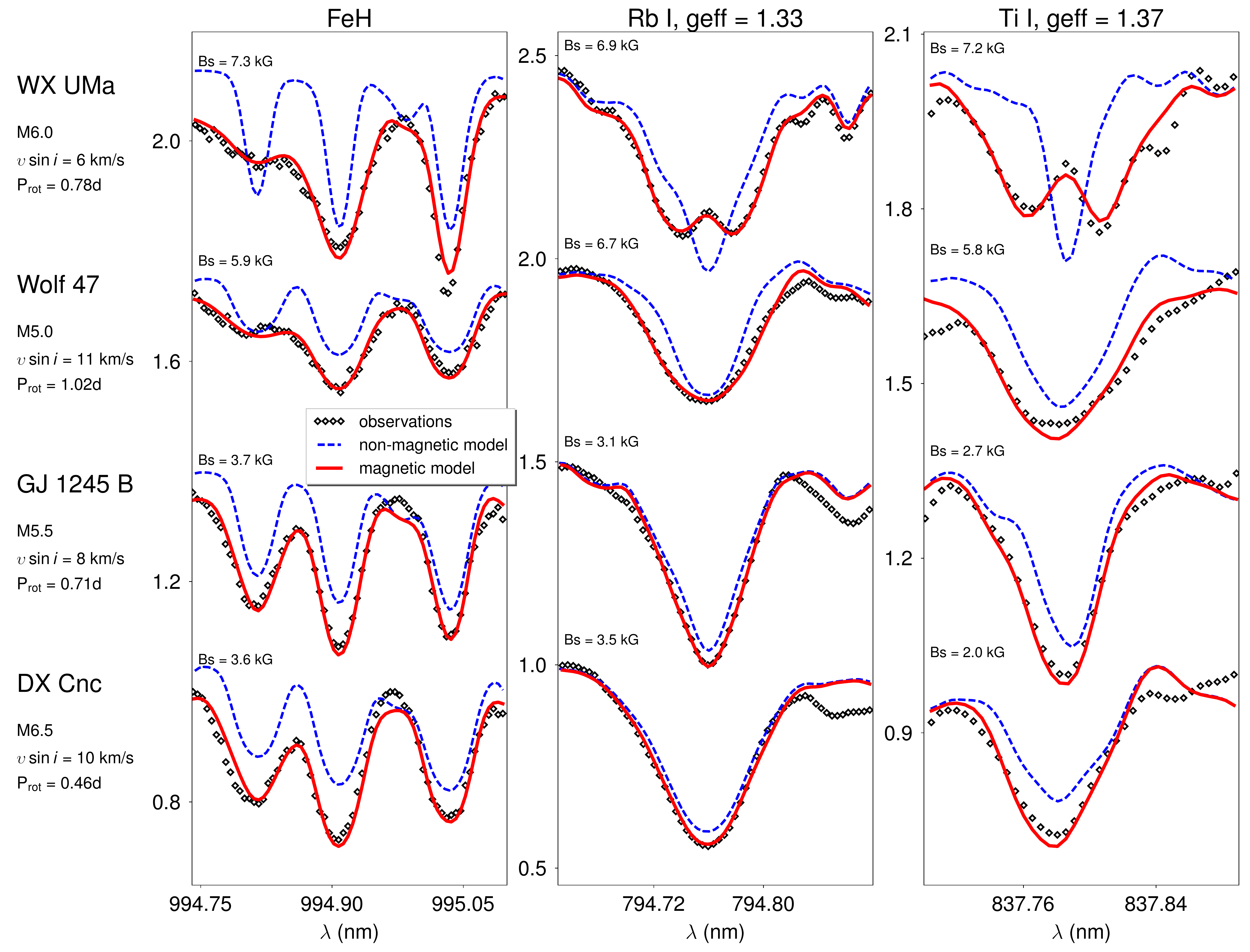}
  }
  \captionof{figure}{
   \label{fig:line-fit}
    \textbf{Magnetic field diagnostics in selected M dwarfs.} 
    \small
    We show example fits to magnetically sensitive spectral lines of FeH molecule (left column), 
    \ion{Rb}{I}~$\lambda794.76$~nm line (middle column), and \ion{Ti}{I}~$\lambda837.7$~nm line (right column)
    in WX~UMa and three other magnetic M dwarfs.
    Black dots~--~observed spectra; blue dashed line~--~predicted zero field model;
    red full line~--~predicted best fit magnetic model.
    The text on the left side lists for each stars its name,
    spectral class, projected rotational velocity $\vsini$, rotational period,
    and surface average magnetic field. Rotation periods are taken from works on
    magnetic mapping\supercite{2010MNRAS.407.2269M}.
    The magnetic field was calculated from distributions of filling factors
    $\bs=\sum_i B_i \cdot f_i$ (see Methods). Our measurements
    of $\bs$ may vary with assumed stellar parameters (e.g., $\teff$, $\logg$) and we show here
    only one choosen solution. In case of WX~UMa, a clear splitting is visible in atomic \ion{Rb}{I} and \ion{Ti}{I} lines,
    and the separation of Zeeman components corresponds to a minimum magnetic field of $\approx6$~kG.
    Employing different $\teff$'s and spectral lines results in magnetic
    fields between $7.0$~kG and $7.5$~kG.}
\end{figure}

\begin{figure}
  \centerline{
    \includegraphics[width=0.7\hsize]{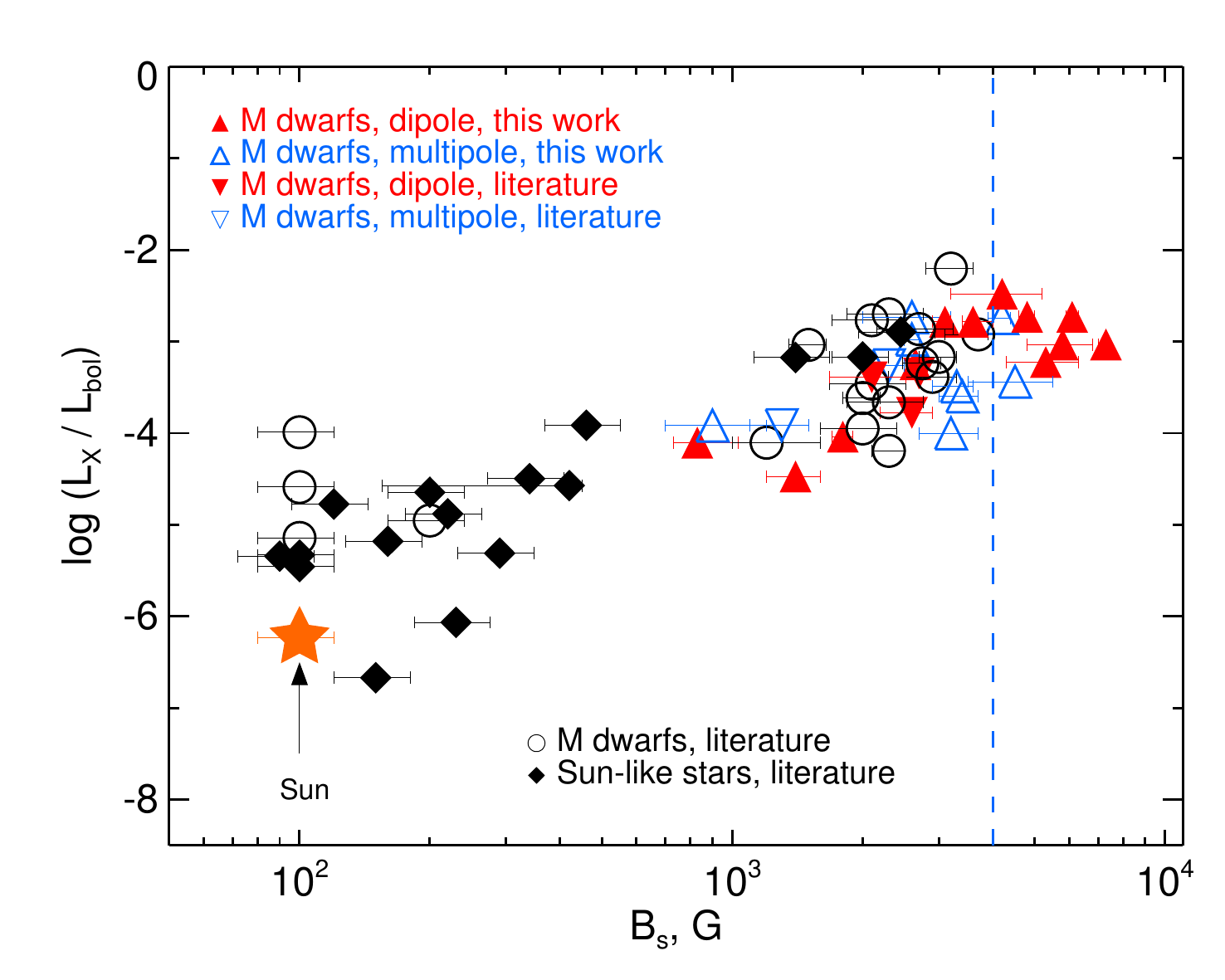}
  }
  \captionof{figure}{
    \label{fig:saturn}
    \textbf{Magnetic field~--~activity relation.} 
    The plot illustrates activity level (in terms of X-ray luminosity, $L_{\rm x}$, normalized to bolometric luminosity $L_{\rm bol}$),
    as a function of magnetic field strength in cool stars.
    Red filled upward and downward triangles are M dwarfs with dipole dominant fields and measurements from this work and literature, respectively.
    Blue open upward and downward triangles are M dwarfs with multipole dominant fields and measurements from this work and literature, respectively.
    Black open circles are M dwarfs with unknown magnetic field configurations and measurements from the literature.
    Sun-like stars are shown with black filled diamonds: we plot these stars because they have
    interior structure similar to partly convective M dwarfs (e.g., both have outer convective zone and inner radiative core)
    and obey same rotation-activity relation\supercite{2014ApJ...794..144R}.
    The Sun is marked with the orange star symbol. 
    Vertical blue dashed line marks the $4$~kG limit of the maximum magnetic field strength measured in early studies.
    It is seen that adding our new measurements
    substantially extends the range of possible mean magnetic fields in M dwarfs.
    The error bars on literature values are taken from original papers (see Supplementary  Table~1),
    while errors on our measurements represent estimated uncertainty rather than formal fitting errors
    (see Fig.~3 and Supplementary Table~2 for more details and notes on individual objects).
  }
\end{figure}

\begin{figure}
  \centerline{
    \includegraphics[width=0.7\hsize]{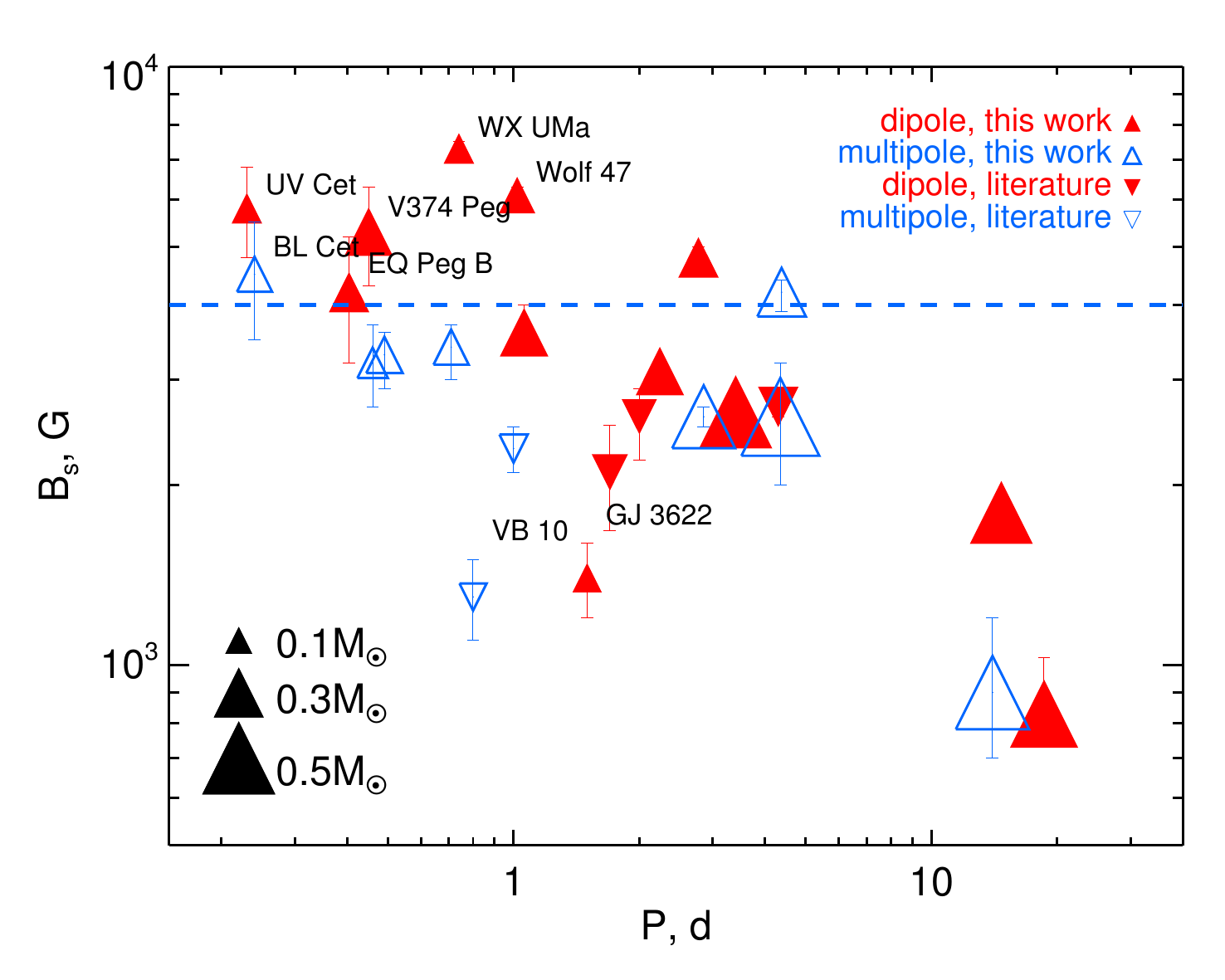}
  }
  \captionof{figure}{
   \label{fig:bf-period}
    \textbf{Rotation~--~magnetic field relation.} 
    \small
    We plot magnetic flux density fields in stars from our sample as a function of rotation period.
    Our new measurements in stars with dipole and multipole states are shown as red filled and blue open triangles, respectively.
    The red and blue upsidedown triangles represent stars with dipole and multipole states, respectively, and magnetic field
    measurements from the literature.
    The symbol size scales with stellar mass (see legend on the plot).
    The horizontal dashed line marks the $4$~kG threshold of saturated magnetic field assigned to stars
    with multipole dynamo regime. We demonstrate that stars with dipole state
    can still generate very strong magnetic fields even at very fast rotation while stars with multipole state
    saturate at rotation periods of about $4$d and fields about $\sim4$~kG. 
    Two stars, VB~10 and GJ~3622, have very weak magnetic fields as expected for their rotation periods.
    We explicitly mark them on the plot and note that it is important to cross check 
    their rotation periods with future observations.
    Measurements from this work that we show on this figure are additionally listed in Supplementary Table~2.
    The error bars on literature values are taken from original papers (see Supplementary Table~1),
    while errors on our measurements represent estimated uncertainty rather than formal fitting errors (see Supplementary Table~2).
    It was calculated as a mean of the four independent measurements (two different effective temperatures and two different spectral regions containing FeH and Ti lines, respectively),
    with error bars being the amplitude of the scatter between these measurements. For stars with largest $\vsini$'s, Bl~Cet, UV~Cet, V374~Peg and EQ~Peg~B, we compute final magnetic field from Ti lines only. 
    We do this because FeH lines become strongly blended and tend to underestimate magnetic fields at cool temperatures
    due to degeneracy between magnetic field and Fe abundance (see section Methods for more details). Considering this and other uncertainties we put a conservative 
    $1$~kG error bars on measured magnetic fields in these stars.
}
\end{figure}

\clearpage

\begin{refsegment}
\defbibfilter{notmethods}{not segment=\therefsegment}

\begin{methods}
\section*{Details on theoretical modelling.}
Our strategy to measure Zeeman splitting is to compare observed
spectra to synthetic spectra computed with a polarized radiative
transfer code that is based on the custom routine taken from the model
atmosphere code \llm\supercite{2004A&A...428..993S} and
extended by the complete treatment of polarized radiation in all four
Stokes parameters\supercite{2006A&A...448.1153K}. The radiative
transfer equation is solved with 
the DELO (Diagonal Element Lambda Operator) method\supercite{1989ApJ...339.1093R,2002A&A...381..736P}, 
which have proved their capabilities for accurate solution of the transfer equation.

In our equation of state we include $99$ atoms (from Hydrogen to Einsteinium) and $524$ molecules.
This results in total of $930$ individual plasma components (i.e., atoms, molecules, and their ions). 
Most of atomic partition functions
are those provided by R.L.~Kurucz\supercite{2005MSAIS...8...14K} (\url{http://kurucz.harvard.edu}).
For molecules we utilize partition functions and equilibrium constants
available from the literature\supercite{1981ApJS...45..621I,1987A&A...182..348I,1984ApJS...56..193S,1988A&AS...74..145I}.
Molecular data can be used in wide temperature ranges from $1000$~K to $9000$~K, i.e., suitable for conditions
found in atmospheres of M dwarf stars.
The final system of linear equations is solved using LU factorization methods available from 
the LAPACK numerical library (\url{http://www.netlib.org/lapack/}).
We tested solutions of our equation of state against those implemented in such codes 
as \textsc{SYNTHE}\supercite{2005MSAIS...8...14K},
\textsc{SSynth}\supercite{1978PhDT.........8I}, 
and \textsc{Synth3}\supercite{2007pms..conf..109K}, and found
good agreement in resulting number densities of most abundant species.

{We treat pressure broadening by including contributions from hydrogen and helium atoms only.
Normally, a contribution from the molecular hydrogen H$_{\rm 2}$ is also expected. However, similar to 
previous studies\supercite{2007ARep...51..282P}, we find that inclusion of H$_{\rm 2}$ overestimates the widths of line
profiles in benchmark inactive M dwarfs. This problem is most prominent in strong alkali lines. 
It increases with decreasing temperature and starts to affect noticeably 
other atomic and molecular lines, too.}

Another source of line broadening is the velocity field caused by convective motions.
However, as shown by 3D hydrodynamical simulations\supercite{2009A&A...508.1429W}, 
the velocity fields in atmospheres of M dwarfs with temperatures $\teff<3500$~K
are well below $1$~\kms. This only has a weak impact on line profiles, leaving the Zeeman effect and rotation
to be the dominating broadening mechanisms. Therefore we assumed zero micro- and macroturbulent velocities
in all calculations.

\section*{Representation of surface magnetic field.}
Our analysis relies on unpolarized \stokesi\ spectra, which means
that we can measure the surface averaged magnetic flux density 
(or simply magnetic field strength) $\bs$
from magnetically broadened line profiles but we derive no information
about field geometry. Note that in case of complex magnetic fields
localized in spots, spot groups, or active regions across the stellar
surface, the mean magnetic field cannot be represented by a simple
homogeneous magnetic field configuration. 
Instead, the best approximation is to assume that the total field 
is the weighted sum over different
field components; $\bs=\btimesfi$, where $f_i$ are corresponding
filling factors that represent the fraction of the stellar surface
covered by the magnetic field of the strength $\b_{i}$. 
We use $11$ magnetic field components and corresponding filling factors are normalized so that $\sum f_i=1$.
In case of stars with large $\vsini$'s individual features in lines profiles 
caused by complex magnetic fields are usually not recoverable and we
speed up computations by assuming {two component model: non-magnetic
and magnetic with a filling factor $f$ so that the total magnetic field strength is $\bs=Bf$.}
In all computations we assumed that the magnetic field is dominated by its
radial component\supercite{2014A&A...563A..35S}.

\section*{Details on spectroscopic analysis.}
In order to compute theoretical line profiles we used MARCS model atmospheres\supercite{2008A&A...486..951G}.
The effective temperatures of M dwarfs are not accurately known and different sources sometimes list
noticeably different values. Therefore, we fit each star assuming two different temperatures. 
{First is the one that follows from spectral types employing
dedicated calibrations\supercite{1995ApJS..101..117K,2004AJ....127.3516G}.
The spectral types were taken from previous works on magnetic field measurements\supercite{2012LRSP....9....1R} 
and from the SIMBAD online service.} 
In our second estimate of the $\teff$ we used lines of TiO $\gamma$-band at $705$~nm 
extracted from Vienna Atomic Line Database (VALD)\supercite{1995A&AS..112..525P,1999A&AS..138..119K} 
with transition parameters calculated by B.~Plez (private communication), version
of 7 Mar 2012, with wavelengths corrected using
polynomial fits to laboratory wavelengths\supercite{1986ApJ...309..449D,1999ApJS..122..331R} (see Master thesis 
by T.~Nordlander\supercite{Nordlander_2012} \url{http://urn.kb.se/resolve?urn=urn:nbn:se:uu:diva-174640}). 
Our temperature estimates from TiO lines are in agreement 
(within error bars) with alternative independent estimates\supercite{2015ApJ...804...64M}.
The other essential parameter~--~stellar gravity ($\logg$)~--~was calculated from the stellar radii
and mass {assuming $R\propto M^{0.9}$ relation which provides surface gravities close to those predicted
by stellar evolution models\supercite{2008ApJS..178...89D,2012MNRAS.427..127B}.}

To derive magnetic fields we used two sets of spectral lines.
First set includes lines of FeH molecule. In particular, we used lines of the Wing-Ford $F^4\,\Delta-X^4\,\Delta$ transitions between $\lamlam990.0-990.4$~nm
and $\lamlam993.8-996.0$~nm regions, respectively.
These lines are a perfect indicators
of magnetic fields in cool stars because of their high 
magnetic sensitivity\supercite{2001ASPC..223.1579V,2006ApJ...644..497R}. 
The transition parameters of FeH molecule were taken from our previous investigations\supercite{2014A&A...563A..35S,2011MNRAS.418.2548S,2010A&A...523A..37S}.

It is known that the Zeeman splitting in FeH lines could not be accurately predicted
by available theoretical descriptions\supercite{2006ApJ...636..548A,2002A&A...385..701B} 
and alternative semi-empirical methods are used instead\supercite{2008A&A...482..387A,2010A&A...523A..37S}. 
Therefore, extending our analysis to atomic lines provides important independent 
constraints on the measured magnetic fields. Therefore, as the second set of magnetic field indicators, 
we choose to use \ion{Ti}{i} lines in $\lamlam960-980$~nm region.
{These lines are originating from the same multiplet and thus their relative oscillator strengths are known with a high precision.
Therefore it was enough to cross check the oscillator strengths
of a few of these lines visible in the spectrum of the Sun\supercite{1985S&T....70...38K} to make sure that we do not
introduce any biases in our analysis.}
These Ti lines have very different magnetic sensitivity with Land\'e g-factors ranging
from $0$ to $1.55$ and different Zeeman patterns 
which makes them very good diagnostics of the magnetic fields.
In addition, we analyzed profiles of \ion{Ti}{i} $\lambda837.7$~nm
and \ion{Rb}{i} $\lambda794.8$~nm lines that show a clear Zeeman splitting
in WX~UMa, but because of very strong blending by TiO molecule in this spectral region 
we could not use these two lines for the final magnetic field measurements {in all stars. 
Instead, we use \ion{Rb}{i} $\lambda794.8$~nm and \ion{Ti}{i} $\lambda837.7$~nm only in Fig.~1 
for illustrative purpose only, i.e., to highlight a very strong magnetic field in WX~UMa. Transition parameters
of all atomic lines were extracted from VALD.}

When the $\vsini$ of a star is large, our measurements rely on the
effect of magnetic intensification of spectral lines\supercite{2004ASSL..307.....L}
which predicts that the depth of a magnetic sensitive line broadened by rotation
will be increased depending on its Zeeman pattern. 
Of course, this
technique is sensitive to fields that are strong enough to produce
observable changes in the equivalent widths of lines 
(see, e.g., Supplementary Fig.~9
for a clear example of magnetic intensification influencing profiles of spectral lines). 
The \ion{Ti}{i} lines in $\lamlam960-980$~nm region are
therefore superior diagnostic of strong fields in fast rotating stars
and we successfully used their magnetic intensification in our analysis.
The reason why these lines are normally completely ignored in spectroscopic
studies is because they are heavily contaminated by telluric
absorption from the Earth atmosphere. Considering the importance of these
lines for our goals we made an effort of removing telluric absorption from all stellar spectra 
of all stars by applying the \textsc{molecfit}
software package\supercite{2015A&A...576A..77S,2015A&A...576A..78K}.

We use the Levenberg-Marquardt minimization algorithm to find
best fit values of filling factors for a given combination of atmospheric parameters\supercite{2014A&A...563A..35S}. 
We simultaneously
treat rotational velocity $\vsini$, atmospheric abundance of a given element,
continuum scaling factor for each spectral region that we fit,
and $11$ filling factors distributed between $0$~kG and $10$~kG as free parameters.
(For stars with very strong field it was sometimes necessary to use wider range of magnetic filling
factors, i.e., $11$ filling factors distributed between $0$~kG and $20$~kG).

\section*{Determination of surface magnetic fields.}
We measured the average magnetic field for each of our sample stars
by fitting synthetic spectra to the data as described above.

We summarize our results in Supplementary Table~2
and show our best fit to profiles of spectral lines 
in Supplementary Figs.~4-22.
Below we give detailed notes on our analysis.

Assuming two different temperature from available photometric calibrations
and from the fit TiO $\gamma$-band, 
we can, in general, obtain similar magnetic field estimates {from Ti and FeH lines 
for at least one of the two temperatures that we tried for each star.} 
However, in some cases a cooler model would result in weaker magnetic field {when derived from FeH lines
and at large $\vsini$ values}. 
This is simple to understand because at large $\vsini$ we can
only measure strong magnetic fields and only through the effect of magnetic intensification of spectral
lines. When we decrease temperature it makes FeH lines stronger, and because most
of FeH lines are magnetic sensitive, it works just the same way as magnetic intensification does.
The code then tries to {adjust} the abundance of Fe to the level when a satisfactory
fit is obtained for all FeH lines while keeping magnetic field low. 
This degeneracy between the temperature and abundance 
on one side, and the magnetic field on another is partly broken for Ti lines
because we have just a few lines, they are well separated from each other,
and one of these lines is completely magnetic insensitive.
All this helps to constrain desired physical parameters from Ti lines more accurately.
{In Gl~182 we measure stronger field from FeH lines compared to Ti lines
which is very likely because of a poor data quality in the region of FeH lines (see Supplementary Fig.~16).}

For GJ~3622 we measure magnetic field which is too weak as for its rotation period 
of about $1.5$d\supercite{2010MNRAS.407.2269M}. Together with another M dwarf
where the measured field seems to be inconsistent with its rotation period, VB~10,
the two lay considerably below the range of magnetic fields that we measured in other 
stars with saturated activity (See Fig.~3). 
We have no explanation why these stars have such weak fields. Note that our estimate
of the magnetic field in GJ~3622 is consistent with previous studies\supercite{2010ApJ...710..924R}.
One reason could be that rotation periods are not accurate: they were estimated from
sparse \stokesv\ spectra for GJ~3622 and simply from stellar $\vsini$ for VB~10, respectively.
Thus, additional photometric monitoring is needed to confirm the periods.

{In EQ~Peg~B we also measure field at the boundary between dipole and multipole state stars ($\bs\approx4$~kG),
but the star has short period and dipole-like magnetic field geometry so that we would expect to see field stronger than this. 
Contrary, in BL~Cet we measure field of $\approx4.5$~kG and its dynamo state is multipole, so one would expect the star to have the field below $4$~kG.
We note again that for all stars with large $\vsini$'s our error bars are also large ($\pm1$~kG) so that these two cases
do not contradict the overall picture. 
Note that EQ~Peg~B was observed only once and thus it is still possible that this was done occasionally
at times when the star's global magnetic field had more simple geometry while the star itself is actually in multipole dynamo state.
It is thus important to monitor this star for a longer period of time in order 
to see any variations in its global magnetic field topology.}

In WX~UMa we clearly observe Zeeman splitting in \ion{Rb}{i}~$\lambda794.76$~nm and
\ion{Ti}{i}~$\lambda837.7$~nm lines. 
From these observed splittings we measure a field of about $6$~kG.
However, this is likely the lower limit for the strength of the field because complex
magnetic fields in M dwarfs tend to produce narrow line cores and wide wings
of magnetic very sensitive spectral lines\supercite{1996IAUS..176..237S,2010A&A...523A..37S,2014A&A...563A..35S}.
When we try to fit FeH lines with our method (i.e., by applying magnetic filling factors as described above)
we find solutions around $7.0$~kG.
Applying the same procedure to Ti lines leads to the average field of about $9.0$~kG for both temperatures that we tried.
However, we observe that strong fields appear because our fitting algorithm tries to fit tiny details 
in line profiles by increasing contribution from
strong magnetic field components when the adjustment of other
free parameters (e.g., abundance) does not help.
Although our combined spectrum of WX~UMa has 
very high signal-to-noise ratio of about $S/N=350$ in the region of Ti lines, there are
additional uncertainties that complicate the fit, e.g., artifacts of telluric removal 
and poor understanding of background  molecular absorption. 
All these can be the reason why we find systematically higher fields from Ti lines compared to what we measure from FeH.
Note that if we ignore this high field components we obtain magnetic field of about $7.3$~kG, i.e. consistent
with measurements from FeH lines.
We believe that a combination of the cool temperature of WX~UMa and its small $\vsini$
make the fit more challenging compared to other stars. 
Indeed, as mentioned above, we did not encounter this problem in stars that rotate
faster than WX~UMa and we can still measure consistent fields from Ti and FeH, respectively. 
Considering all the uncertainties in our modelling of this star, we trust
results from FeH lines most.

\section*{Mass dependence of stellar magnetic fields.}
Our measurements show that stars with simple magnetic field configurations can generate strongest
magnetic fields. However, these objects can have very different masses from $0.1\Msun$ of WX~UMa
to $0.28\Msun$ of V374~Peg. In Supplementary Fig.~2 we plot our magnetic field measurements
as a function of Rossby number $Ro=P/\tau_{\rm c}$, where $P$ is the rotation period and
$\tau_{\rm c}$ is the convective turnover time. We computed turnover times using commonly adopted empirical
calibrations \supercite{2011ApJ...743...48W}. The symbol size on this plot scales with stellar mass.
As expected, there are no additional trends seen from this plot compared to what is shown in our Fig.~3. 
Both partly ($M>0.35\Msun$) and fully convective ($M<0.35\Msun$) stars can generate fields of about few kilogauss.
This result is in agreement with previous works from the analysis of \stokesv\supercite{2010MNRAS.407.2269M}.
Note that all stars where we measure strong fields
belong to the saturated activity branch (i.e., $\log Ro<-1$)
where the activity level is believed to be independent on Rossby number\supercite{1984ApJ...279..763N,2003A&A...397..147P,2011ApJ...743...48W,2014ApJ...794..144R}.

\section*{Comparison with previous measurements.}
In this work we used method of filling factors to derive magnetic field strength from unpolarized light.
Our measurements are in a good agreement with previously published results\supercite{2012LRSP....9....1R} within 
commonly adopted error bars of about $1$~kG\supercite{2007ApJ...656.1121R}. However, for GJ~1245~B and DX~Cnc our measurements from FeH lines
are about $1.7$~kG larger than that quoted before\supercite{2007ApJ...656.1121R}. The possible reason can be that here we use direct
radiative transfer modelling against a template interpolation which was used before. It also can be that we underestimate
pressure broadening of FeH lines which leads to the higher fields especially at coolest temperatures, which is expected.
Indeed, by including broadening by molecular hydrogen we can obtain noticeably lower fields of $3.1$~kG for GJ~1245~B
and $2.7$~kG for DX~Cnc, respectively. 
Note that these values are still about $1$~kG stronger than previously estimated. We also checked the effect of including
broadening by molecular hydrogen in our computations for other stars and found no substantial deviations from values that we present in this work.
The most affected parameters are the abundances of Ti and Fe (with additional broadening abundances tend to be reduced, as expected), 
while rotational velocities and magnetic field strength remain mostly consistent between the two calculations.
For YZ~CMi different studies report very different fields from $3.3$~kG\supercite{2000ASPC..198..371J} to $4.5$~kG\supercite{2009AIPC.1094..124K}.
Our current estimate is $4.8$~kG, and it is significantly higher compared to $\bs=3.6$~kG that we measured in our previous work\supercite{2014A&A...563A..35S}.
This is because that time we derived stellar rotational velocity and iron abundance separately from filling factors by a simple
manual match which led to a weaker fields measured.
Note that we obtain consistent field measurements in this star independent on assumed H$_{\rm 2}$ broadening,
so that differences between literature values are most likely due to different methods used.
For WX~UMa there are no magnetic field measurements available, only a lower limit of $B>3.9$~kG was set because of limitations of a method used in previous works\supercite{2009ApJ...692..538R}.
For the binary system Gl~65~AB we measure field of about $4.5\pm1.0$~kG in the primary and $5.8\pm1.0$~kG in the secondary, respectively. Both these values are systematically lower
compared to $5.2\pm0.5$~kG and $6.7\pm0.6$~kG reported in Kochukhov \& Lavail (2017)\supercite{2017ApJ...835L...4K}.
This discrepancy results from the difference in fitting methods utilized in both works. However, note that stars with large $vsini$'s
are still subject to large fitting uncertainties in both approaches. From the direct comparison of Zeeman-sensitive lines we see that Gl~65~B has clearly stronger field
compared to Gl~65~A since both stars have close spectral types. Considering that the magnetic field in Gl~65~A has complex multipole geometry 
and in Gl~65~B has more simple dipole-like geometry\supercite{2017ApJ...835L...4K},
our magnetic measurements for these two stars agree with the rest of our conclusions.

\section*{Data availability}
The data that support the plots within this paper and other findings of this study are available from the corresponding author upon reasonable request.

\end{methods}
\printbibliography[segment=\therefsegment,resetnumbers=false,filter=notmain]
\end{refsegment}

\begin{supplementaryinformation}
\begin{refsegment}
\defbibfilter{notsupp}{not segment=\therefsegment}

\renewcommand{\tablename}{\sffamily\noindent\textbf{Supplementary Table}}
\sffamily\scriptsize
\begin{longtable}[H]{llccccc}

\caption{Data used to create Fig.~\ref{fig:saturn}.}\label{tab:saturn}\\
\hline
                       &                         &                         &                        &              &                                                    &              \\
Star name              &  Alternative name       & Spectral type           & $\bs$                  &   Reference  & $\displaystyle\log\frac{L_{\rm X}}{L_{\rm bol}}$   & Dynamo state \\
                       &                         &                         & G                      &              &                                                    &              \\
                       &                         &                         &                        &              &                                                    &              \\
\hline
\endfirsthead                                                                                                                                                           

\multicolumn{7}{c}{\tablename\ \thetable\ -- \textit{Continued from previous page}} \\                                                                                                                                                                      

\hline                                                                                                                                                                  
                       &                         &                         &                        &              &                                                    &              \\
Star name              &  Alternative name       & Spectral type           & $\bs$                  &   Reference  & $\displaystyle\log\frac{L_{\rm X}}{L_{\rm bol}}$   & Dynamo state \\
                       &                         &                         & G                      &              &                                                    &              \\
                       &                         &                         &                        &              &                                                    &              \\
\hline 
\endhead

\hline 
\multicolumn{7}{r}{\textit{Continued on next page}} \\
\hline 
\endfoot

\hline
\endlastfoot

Gl 504                   & 59 Vir         & G0.0 &     $420^{+30}_{-265}$   &   \cite{2010A&A...522A..81A} & -4.57 &  \\ 
Gl 506                   & 61 Vir         & G6.0 &     $150^{+30}_{-30}$    &   \cite{2010A&A...522A..81A} & -6.67 &  \\ 
Gl 566 A                 & $\xi$ Boo A    & G8.0 &     $340^{+68}_{-68}$    &   \cite{1994ASPC...64..438L} & -4.50 &  \\ 
Gl 702 A                 & 70 Oph A       & K0.0 &     $220^{+44}_{-44}$    &   \cite{1989ApJ...345..480M} & -4.88 &  \\ 
Gl 764                   & $\sigma$ Dra   & K0.0 &     $100^{+20}_{-20}$    &   \cite{1995ApJ...439..939V} & -5.33 &  \\ 
Gl 166 A                 & 40 Eri         & K1.0 &     $100^{+20}_{-20}$    &   \cite{1995ApJ...439..939V} & -5.46 &  \\ 
Gl 663 B                 & 36 Oph B       & K1.0 &     $120^{+24}_{-24}$    &   \cite{1997A&A...318..429R} & -4.77 &  \\ 
Gl 355                   & LQ Hya         & K2.0 &     $2450^{+490}_{-490}$ &   \cite{1996IAUS..176..237S} & -2.90 &  \\ 
Gl 663 A                 & 36 Oph A       & K2.0 &     $200^{+40}_{-40}$    &   \cite{1989ApJ...345..480M} & -4.65 &  \\ 
Gl 706                   &                & K2.0 &     $230^{+46}_{-46}$    &   \cite{1988ApJ...330..274B} & -6.07 &  \\ 
Gl 566 B                 & $\xi$ Boo B    & K4.0 &     $460^{+92}_{-92}$    &   \cite{1994IAUS..154..493S} & -3.92 &  \\ 
Gl 570 A                 &                & K4.0 &     $160^{+32}_{-32}$    &   \cite{1997A&A...318..429R} & -5.18 &  \\ 
Gl 517                   & EQ Vir         & K5.0 &     $2000^{+300}_{-300}$ &   \cite{1986ApJ...302..777S} & -3.17 &  \\ 
Gl 171.2 A               &                & K5.0 &     $1400^{+280}_{-280}$ &   \cite{1996IAUS..176..237S} & -3.17 &  \\ 
Gl 820 A                 &61 Cyg A        & K5.0 &     $290^{+58}_{-58}$    &   \cite{1989ApJ...345..480M} & -5.31 &  \\ 
Gl 845                   &$\epsilon$ Ind  & K5.0 &      $90^{+18}_{-18}$    &   \cite{1997A&A...318..429R} & -5.34 &  \\ 
Gl 182                   &                & M0.0 &    $2600^{+600}_{-600}$  &   this work                  & -2.73 &   multipole \\ 
Gl 410                   & DS Leo         & M1.0 &    $900^{+300}_{-200}$   &   this work                  & -3.74 &     multipole  \\ 
Gl 803                   & AU Mic         & M1.0 &    $2300^{+460}_{-460}$  &   \cite{1994IAUS..154..493S} & -2.70 &   \\ 
GJ 9520                  & OT Ser         & M1.0 &    $2700^{+100}_{-100}$  &   this work                  & -3.25 &   dipole   \\ 
Gl 49                    &                & M1.5 &     $800^{+200}_{-100}$  &   this work                  & -4.10 &   dipole   \\ 
Gl 70                    &                & M2.0 &     $100^{+20}_{-20}$    &   \cite{2007ApJ...656.1121R} & -4.59 &    \\ 
Gl 494                   & DT Vir         & M2.0 &    $2600^{+100}_{-100}$  &   this work                  & -2.98 &     multipole  \\ 
Gl 569 A                 & CE Boo         & M2.0 &    $1800^{+100}_{-100}$  &   this work                  & -4.04 &     dipole  \\ 
Gl 388                   & AD Leo         & M3.5 &    $3100^{+100}_{-200}$  &   this work                  & -2.78 &     dipole   \\ 
Gl 729                   &                & M3.5 &    $2300^{+200}_{-200}$  &   \cite{2014A&A...563A..35S} & -4.19 &    \\ 
Gl 873                   & EV Lac         & M3.5 &    $4200^{+200}_{-300}$  &   this work                  & -2.74 &     multipole  \\ 
Gl 896 A                 & EQ Peg A       & M3.5 &    $3600^{+400}_{-200}$  &   this work                  & -2.78 &    dipole   \\ 
GJ 3379                  &                & M3.5 &    $2300^{+460}_{-460}$  &   \cite{2009ApJ...692..538R} & -3.66 &  \\ 
GJ 4247                  & V374 Peg       & M3.5 &    $5300^{+1000}_{-1000}$&   this work                  & -3.08 &    dipole   \\ 
Gl 490 B                 &                & M4.0 &    $3200^{+400}_{-400}$  &   \cite{2009ApJ...704.1721P} & -2.20 &    \\ 
Gl 852 A                 &                & M4.0 &    $3000^{+300}_{-300}$  &   \cite{2011MNRAS.418.2548S} & -3.17 &   \\ 
Gl 876                   &                & M4.0 &     $200^{+40}_{-40}$    &   \cite{2007ApJ...656.1121R} & -4.95 &    \\ 
GJ 1005 A                &                & M4.0 &     $100^{+20}_{-20}$    &   \cite{2007ApJ...656.1121R} & -5.14 &    \\ 
GJ 2069 B                &                & M4.0 &    $2700^{+540}_{-540}$  &   \cite{2009ApJ...692..538R} & -2.86 &    \\ 
Gl 234 A                 &                & M4.5 &    $2750^{+275}_{-275}$  &   \cite{2011MNRAS.418.2548S} & -3.24 &  \\ 
Gl 285                   & YZ CMi         & M4.5 &    $4800^{+200}_{-200}$  &   this work                  & -2.73 &       dipole  \\ 
Gl 493.1                 &                & M4.5 &    $2100^{+420}_{-420}$  &   \cite{2009ApJ...692..538R} & -3.46 &  \\ 
Gl 852 B                 &                & M4.5 &    $1500^{+150}_{-150}$  &   \cite{2011MNRAS.418.2548S} & -3.04 &   \\ 
Gl 896 B                 & EQ Peg B       & M4.5 &    $4200^{+1000}_{-1000}$&   this work                  & -2.48 &     dipole   \\ 
GJ 1224                  &                & M4.5 &    $2700^{+100}_{-100}$  &   \cite{2007ApJ...656.1121R} & -3.35 &   dipole$^*$ \\ 
GJ 4053                  &                & M4.5 &    $2000^{+400}_{-400}$  &   \cite{2009ApJ...692..538R} & -3.95 &    \\ 
Gl 51                    & Wolf 47        & M5.0 &    $6100^{+200}_{-200}$  &   this work                  & -2.71 &    dipole  \\ 
Gl 905                   &                & M5.0 &     $100^{+20}_{-20}$    &   \cite{2007ApJ...656.1121R} & -3.99 &    \\ 
GJ 1154 A                &                & M5.0 &    $2100^{+420}_{-420}$  &   \cite{2009ApJ...692..538R} & -3.39 &  dipole$^*$  \\ 
GJ 1156                  & GL Vir         & M5.0 &    $3300^{+300}_{-400}$  &   this work                  & -3.49 &   multipole  \\ 
Gl 65 A                  & BL Cet         & M5.5 &    $4500^{+1000}_{-1000}$&   this work                  & -3.44 &  multipole \\ 
Gl 406                   &                & M5.5 &    $2600^{+300}_{-400}$  &   \cite{2014A&A...563A..35S} & -3.78 &  dipole$^*$ \\ 
GJ 1002                  &                & M5.5 &       $0^{+0}_{+0}$      &   \cite{2007ApJ...656.1121R} & -5.45 &    \\ 
GJ 1245 B                &                & M5.5 &    $3400^{+300}_{-400}$  &   this work                  & -3.60 &   multipole \\ 
GJ 2005 A                &                & M5.5 &    $2000^{+200}_{-200}$  &   \cite{2011MNRAS.418.2548S} & -3.62 &   \\ 
Gl 65 B                  & UV Cet         & M6.0 &    $5800^{+1000}_{-1000}$&   this work                  & -3.03 &  dipole \\ 
Gl 412 B                 & WX UMa         & M6.0 &    $7300^{+200}_{-300}$  &   this work                  & -3.03 &   dipole \\ 
GJ 1111                  & DX Cnc         & M6.5 &    $3200^{+500}_{-500}$  &   this work                  & -4.00 &   multipole \\ 
GJ 3622                  &                & M6.5 &    $1400^{+200}_{-200}$  &   this work                  & -4.33 &   dipole  \\ 
Gl 644 C                 & VB 8           & M7.0 &    $2300^{+200}_{-200}$  &   \cite{2007ApJ...656.1121R} & -3.26 &   multipole$^*$ \\ 
2MASS J14563831-2809473  &                & M7.0 &    $1200^{+400}_{-200}$  &   \cite{2010ApJ...710..924R} & -4.10 &    \\ 
Gl 752 B                 & VB 10          & M8.0 &    $1300^{+200}_{-200}$  &   \cite{2007ApJ...656.1121R} & -3.91 &   multipole$^*$ \\ 
2MASS J03205965+1854233  & LP 412-31      & M8.0 &    $3700^{+200}_{-600}$  &   \cite{2010ApJ...710..924R} & -2.92 &  \\ 
2MASS J00244419-2708242B &                & M8.5 &    $2100^{+400}_{-400}$  &   \cite{2010ApJ...710..924R} & -2.77 &  \\ 
2MASS J08533619-0329321  &                & M9.0 &    $2900^{+400}_{-600}$  &   \cite{2010ApJ...710..924R} & -3.39 &    \\ 
Sun                      &                & G2.0 &     $100^{+20}_{-20}$    &   \cite{2009ASPC..405..135S} & -6.23 &   \\ 
\end{longtable}
\vspace{-5ex}
\sffamily\scriptsize
For each star the table lists its name, spectral type, surface magnetic field,
corresponding reference to original publication for the magnetic field, ratio of the X-ray to bolometric luminosity,
and the dynamo state based on published magnetic maps. Dynamo states marked with asterics (*) mean that no ZDI was performed for these stars
and their dynamo states were inferred from the presence or absence of persistent, rotationally-invariant \stokesv\ signature.
The X-ray luminosities were extracted from the NEXXUS database\supercite{2004A&A...417..651S} \textbf{with no error bars provided}.
The error bars on the magnetic field measurements are taken from original works cited in the fifth column.
The error bars on the magnetic field measurements from this work represent estimated uncertainty rather than formal fitting errors (see Supplementary Table~\ref{tab:magnetic}).

\begin{table}[!ht]

\renewcommand{\tablename}{\sffamily\noindent\textbf{Supplementary Table}}

\caption{Magnetic field measurements.\label{tab:magnetic}}
\sffamily\scriptsize
\begin{center}
\begin{tabular}{llcc|ccc|ccc|cc}
\hline
                                                           &                         &                        &                 &                &         &                 &               &         &                                             &                      \\
Star                       &  Alternative name             & $\teff$                 & $\logg$                &  \multicolumn{3}{c|}{\ion{Ti}{i} lines}    &  \multicolumn{3}{c|}{FeH lines}           &                                             & $P$                  \\
                           &                               & K                       & dex                    &   $\abn{Ti}$    & $\vsini$       &  $\bs$  &   $\abn{Fe}$    & $\vsini$      &  $\bs$  & $\bs$                                       & days                 \\
                           &                               &                         &                        &   dex           &  \kms          &   kG    &   dex           &  \kms         &   kG    &  kG                                         &                      \\
\hline
\multirow{2}{*}{Gl 65 B}   &  \multirow{2}{*}{UV Cet}      & 2900                    & 5.2                    & -7.17$\pm$0.01  & 31.4$\pm$0.3   & 5.7     &  -4.35$\pm$0.02 & 32.3$\pm$0.4  &  5.6    & \multirow{2}{*}{$5.8^{+1.0}_{-1.0}$}      &\multirow{2}{*}{0.23} \\
                           &                               & 2800                    & 5.2                    & -7.35$\pm$0.01  & 31.5$\pm$0.3   & 5.9     &  -4.64$\pm$0.02 & 32.7$\pm$0.3  &  5.1    &                                           &                      \\
\hline
\multirow{2}{*}{Gl 65 A}   &  \multirow{2}{*}{BL Cet}      & 3000                    & 5.2                    & -7.19$\pm$0.01  & 29.1$\pm$0.3   & 4.3     &  -4.19$\pm$0.02 & 29.1$\pm$0.3  &  3.6    & \multirow{2}{*}{$4.5^{+1.0}_{-1.0}$}      &\multirow{2}{*}{0.24} \\
                           &                               & 2900                    & 5.2                    & -7.36$\pm$0.01  & 28.9$\pm$0.3   & 4.7     &  -4.48$\pm$0.02 & 29.4$\pm$0.3  &  3.0    &                                           &                      \\
\hline
\multirow{2}{*}{Gl 896 B}  &  \multirow{2}{*}{EQ Peg B}    & 3300                    & 5.0                    & -6.92$\pm$0.02  & 26.2$\pm$0.3   & 4.2     &  -4.00$\pm$0.02 & 26.3$\pm$0.3  &  3.6    & \multirow{2}{*}{$4.2^{+1.0}_{-1.0}$}      &\multirow{2}{*}{0.41} \\
                           &                               & 3100                    & 5.1                    & -7.10$\pm$0.02  & 26.2$\pm$0.3   & 4.1     &  -4.41$\pm$0.02 & 26.6$\pm$0.3  &  3.1    &                                           &                      \\
\hline
\multirow{2}{*}{GJ 4247}   &  \multirow{2}{*}{V374 Peg}    & 3400                    & 5.0                    & -6.96$\pm$0.01  & 35.4$\pm$0.2   & 5.2     &  -3.96$\pm$0.02 & 37.0$\pm$0.2  &  4.8    & \multirow{2}{*}{$5.3^{+1.0}_{-1.0}$}      &\multirow{2}{*}{0.45} \\
                           &                               & 3200                    & 5.1                    & -7.10$\pm$0.01  & 35.3$\pm$0.2   & 5.3     &  -4.34$\pm$0.02 & 37.4$\pm$0.5  &  4.3    &                                           &                      \\
\hline
\multirow{2}{*}{GJ 1111}   &  \multirow{2}{*}{DX Cnc}      & 2900                    & 5.2                    & -6.98$\pm$0.02  & 11.4$\pm$0.3   & 2.7     &  -4.52$\pm$0.03 & 10.2$\pm$0.2  &  3.7    & \multirow{2}{*}{$3.2^{+0.5}_{-0.5}$}      &\multirow{2}{*}{0.46} \\
                           &                               & 2800                    & 5.2                    & -7.13$\pm$0.02  & 11.5$\pm$0.3   & 2.8     &  -4.77$\pm$0.03 & 10.2$\pm$0.2  &  3.6    &                                           &                      \\
\hline
\multirow{2}{*}{GJ 1156}   &  \multirow{2}{*}{Gl Vir}      & 3200                    & 5.1                    & -6.86$\pm$0.02  & 17.2$\pm$0.3   & 2.9     &  -4.30$\pm$0.03 & 15.7$\pm$0.3  &  3.6    & \multirow{2}{*}{$3.3^{+0.3}_{-0.4}$}      &\multirow{2}{*}{0.49} \\
                           &                               & 3000                    & 5.2                    & -7.12$\pm$0.02  & 17.1$\pm$0.3   & 3.6     &  -4.68$\pm$0.03 & 16.1$\pm$0.3  &  2.9    &                                           &                      \\
\hline
\multirow{2}{*}{GJ 1245 B} &  \multirow{2}{*}{}            & 3100                    & 5.1                    & -6.83$\pm$0.03  &  8.0$\pm$0.2   & 3.0     &  -4.24$\pm$0.03 &  7.8$\pm$0.3  &  3.7    & \multirow{2}{*}{$3.4^{+0.3}_{-0.4}$}      &\multirow{2}{*}{0.71} \\
                           &                               & 2900                    & 5.2                    & -7.13$\pm$0.02  &  8.1$\pm$0.2   & 3.3     &  -4.74$\pm$0.03 &  7.9$\pm$0.3  &  3.5    &                                           &                      \\
\hline
\multirow{2}{*}{Gl 412 B}  &  \multirow{2}{*}{WX UMa}      & 3100                    & 5.1                    & -6.64$\pm$0.03  &  5.8$\pm$0.3   & 7.2     &  -4.30$\pm$0.03 &  5.4$\pm$0.4  &  7.3    & \multirow{2}{*}{$7.3^{+0.2}_{-0.3}$}      &\multirow{2}{*}{0.78} \\
                           &                               & 2900                    & 5.2                    & -6.91$\pm$0.02  &  5.8$\pm$0.3   & 7.5     &  -4.79$\pm$0.02 &  5.4$\pm$0.3  &  7.0    &                                           &                      \\
\hline
\multirow{2}{*}{Gl 51}     &  \multirow{2}{*}{Wolf 47}     & 3200                    & 5.1                    & -6.98$\pm$0.02  & 12.4$\pm$0.3   & 6.2     &  -4.50$\pm$0.03 & 10.2$\pm$0.4  &  5.9    & \multirow{2}{*}{$6.1^{+0.2}_{-0.2}$}      &\multirow{2}{*}{1.02} \\
                           &                               & 3000                    & 5.2                    & -7.15$\pm$0.02  & 12.9$\pm$0.2   & 6.3     &  -4.88$\pm$0.03 & 10.3$\pm$0.4  &  5.8    &                                           &                      \\
\hline
\multirow{2}{*}{Gl 896 A}  &  \multirow{2}{*}{EQ Peg A}    & 3400                    & 5.0                    & -6.91$\pm$0.01  & 15.7$\pm$0.2   & 3.6     &  -4.49$\pm$0.04 & 15.2$\pm$0.5  &  3.6    & \multirow{2}{*}{$3.6^{+0.4}_{-0.2}$}      &\multirow{2}{*}{1.06} \\
                           &                               & 3200                    & 5.1                    & -7.10$\pm$0.01  & 15.6$\pm$0.2   & 4.0     &  -4.78$\pm$0.05 & 15.3$\pm$0.5  &  3.2    &                                           &                      \\
\hline
\multirow{2}{*}{GJ 3622}   &  \multirow{2}{*}{}            & 2800                    & 5.2                    & -7.17$\pm$0.03  &  4.6$\pm$0.3   & 1.5     &  -4.61$\pm$0.03 &  3.7$\pm$0.3  &  1.3    & \multirow{2}{*}{$1.4^{+0.2}_{-0.2}$}      &\multirow{2}{*}{1.50} \\
                           &                               & 2700                    & 5.3                    & -7.40$\pm$0.03  &  4.5$\pm$0.3   & 1.6     &  -4.87$\pm$0.03 &  3.9$\pm$0.2  &  1.2    &                                           &                      \\
\hline
\multirow{2}{*}{Gl 388}    &  \multirow{2}{*}{AD Leo}      & 3500                    & 4.9                    & -6.91$\pm$0.01  &  3.4$\pm$0.2   & 2.9     &  -4.53$\pm$0.02 &  2.9$\pm$0.3  &  3.2    & \multirow{2}{*}{$3.1^{+0.1}_{-0.2}$}      &\multirow{2}{*}{2.24} \\
                           &                               & 3300                    & 5.0                    & -7.00$\pm$0.01  &  3.5$\pm$0.2   & 3.0     &  -4.83$\pm$0.02 &  2.9$\pm$0.4  &  3.1    &                                           &                      \\
\hline
\multirow{2}{*}{Gl 285}    &  \multirow{2}{*}{YZ CMi}      & 3400                    & 5.0                    & -6.90$\pm$0.04  &  6.3$\pm$0.4   & 4.5     &  -4.34$\pm$0.03 &  5.2$\pm$0.3  &  4.9    & \multirow{2}{*}{$4.8^{+0.2}_{-0.2}$}      &\multirow{2}{*}{2.77} \\
                           &                               & 3200                    & 5.1                    & -7.10$\pm$0.03  &  5.8$\pm$0.4   & 4.9     &  -4.68$\pm$0.03 &  5.1$\pm$0.3  &  4.9    &                                           &                      \\
\hline
\multirow{2}{*}{Gl 494}    &  \multirow{2}{*}{DT Vir}      & 3800                    & 4.7                    & -6.97$\pm$0.01  & 10.3$\pm$0.1   & 2.7     &  -4.55$\pm$0.03 & 10.8$\pm$0.4  &  2.5    & \multirow{2}{*}{$2.6^{+0.1}_{-0.1}$}      &\multirow{2}{*}{2.85} \\
                           &                               & 3600                    & 4.8                    & -6.97$\pm$0.01  & 10.2$\pm$0.3   & 2.6     &  -4.71$\pm$0.02 & 10.4$\pm$0.4  &  2.5    &                                           &                      \\
\hline
\multirow{2}{*}{GJ 9520}   &  \multirow{2}{*}{OT Ser}      & 3800                    & 4.7                    & -6.91$\pm$0.02  &  4.8$\pm$0.2   & 2.7     &  -4.48$\pm$0.02 &  4.6$\pm$0.3  &  2.7    & \multirow{2}{*}{$2.7^{+0.1}_{-0.1}$}      &\multirow{2}{*}{3.40} \\
                           &                               & 3600                    & 4.8                    & -6.91$\pm$0.02  &  4.9$\pm$0.2   & 2.6     &  -4.63$\pm$0.02 &  4.5$\pm$0.3  &  2.6    &                                           &                      \\
\hline
\multirow{2}{*}{Gl 182}    &  \multirow{2}{*}{}            & 4000                    & 4.6                    & -6.94$\pm$0.02  &  9.2$\pm$0.1   & 2.1     &  -4.51$\pm$0.03 &  9.1$\pm$0.7  &  3.1    & \multirow{2}{*}{$2.6^{+0.6}_{-0.6}$}      &\multirow{2}{*}{4.35} \\
                           &                               & 3800                    & 4.7                    & -6.92$\pm$0.02  &  9.0$\pm$0.1   & 2.0     &  -4.62$\pm$0.05 &  8.8$\pm$0.8  &  3.2    &                                           &                      \\
\hline
\multirow{2}{*}{Gl 873}    &  \multirow{2}{*}{EV Lac}      & 3400                    & 5.0                    & -6.88$\pm$0.02  &  5.4$\pm$0.2   & 4.0     &  -4.52$\pm$0.02 &  4.2$\pm$0.3  &  4.4    & \multirow{2}{*}{$4.2^{+0.2}_{-0.3}$}      &\multirow{2}{*}{4.38} \\
                           &                               & 3200                    & 5.1                    & -7.02$\pm$0.02  &  5.5$\pm$0.2   & 4.1     &  -4.85$\pm$0.02 &  4.2$\pm$0.3  &  4.3    &                                           &                      \\
\hline
\multirow{2}{*}{Gl 410}    &  \multirow{2}{*}{DS Leo}      & 3800                    & 4.7                    & -6.96$\pm$0.02  &  3.2$\pm$0.2   & 0.8     &  -4.72$\pm$0.02 &  2.9$\pm$0.4  &  0.8    & \multirow{2}{*}{$0.9^{+0.3}_{-0.2}$}      &\multirow{2}{*}{14.0} \\
                           &                               & 3600                    & 4.8                    & -6.91$\pm$0.02  &  0.5$\pm$1.0   & 0.7     &  -4.91$\pm$0.02 &  0.4$\pm$1.0  &  1.2    &                                           &                      \\
\hline
\multirow{2}{*}{Gl 569 A}  &  \multirow{2}{*}{CE Boo}      & 3600                    & 4.8                    & -6.95$\pm$0.01  &  2.2$\pm$0.2   & 1.8     &  -4.65$\pm$0.02 &  1.7$\pm$0.6  &  1.8    & \multirow{2}{*}{$1.8^{+0.1}_{-0.1}$}      &\multirow{2}{*}{14.7} \\
                           &                               & 3500                    & 4.9                    & -6.97$\pm$0.01  &  2.5$\pm$0.2   & 1.7     &  -4.76$\pm$0.02 &  1.5$\pm$0.7  &  1.8    &                                           &                      \\
\hline
\multirow{2}{*}{Gl 49}     &  \multirow{2}{*}{}            & 3800                    & 4.7                    & -6.99$\pm$0.04  &  0.1$\pm$0.8   & 0.8     &  -4.58$\pm$0.02 &  0.1$\pm$0.8  &  0.8    & \multirow{2}{*}{$0.8^{+0.2}_{-0.1}$}      &\multirow{2}{*}{18.6} \\
                           &                               & 3600                    & 4.8                    & -6.99$\pm$0.05  &  0.1$\pm$0.9   & 1.0     &  -4.72$\pm$0.02 &  0.1$\pm$1.0  &  0.7    &                                           &                      \\
\hline
\end{tabular}                                                                                                                                                                                               
\end{center}                                                                                                                                                                                                
\sffamily\scriptsize
For each star we list magnetic field measurements and values of derived abundance and rotation velocities 
from Ti and FeH lines, respectively. The results are given for the two assumed effective temperatures that we derived
by using photometric calibrations\supercite{2004AJ....127.3516G,1995ApJS..101..117K} 
and by fitting the strength of the TiO $\gamma$-band at $750$~nm. The finally adopted magnetic field strength is listed in the 11th column.
It was calculated as a mean of the four independent measurements,
with error bars being the amplitude of the scatter between these measurements. {For stars with largest $\vsini$'s, Bl~Cet, UV~Cet, V374~Peg and EQ~Peg~B,
we compute final magnetic field from Ti lines only. We do this because FeH lines become strongly blended and tend to underestimate magnetic fields at cool temperatures
due to degeneracy between magnetic field and Fe abundance (see section Methods for more details). Considering this and other uncertainties we put a conservative 
$1$~kG error bars on measured magnetic fields in these stars.}
The rotation periods were taken from original papers on Zeeman Doppler Imaging\supercite{2010MNRAS.407.2269M,2017ApJ...835L...4K}.
\end{table}

\renewcommand\thefigure{}

\begin{figure}
  \centerline{$i=0$\degr \hspace{4.5cm} $i=90$\degr}
  \centerline{$\bs=1.95$~kG \hspace{2.5cm} $\bs=1.37$~kG}
  \centerline{
   \includegraphics[width=0.3\hsize]{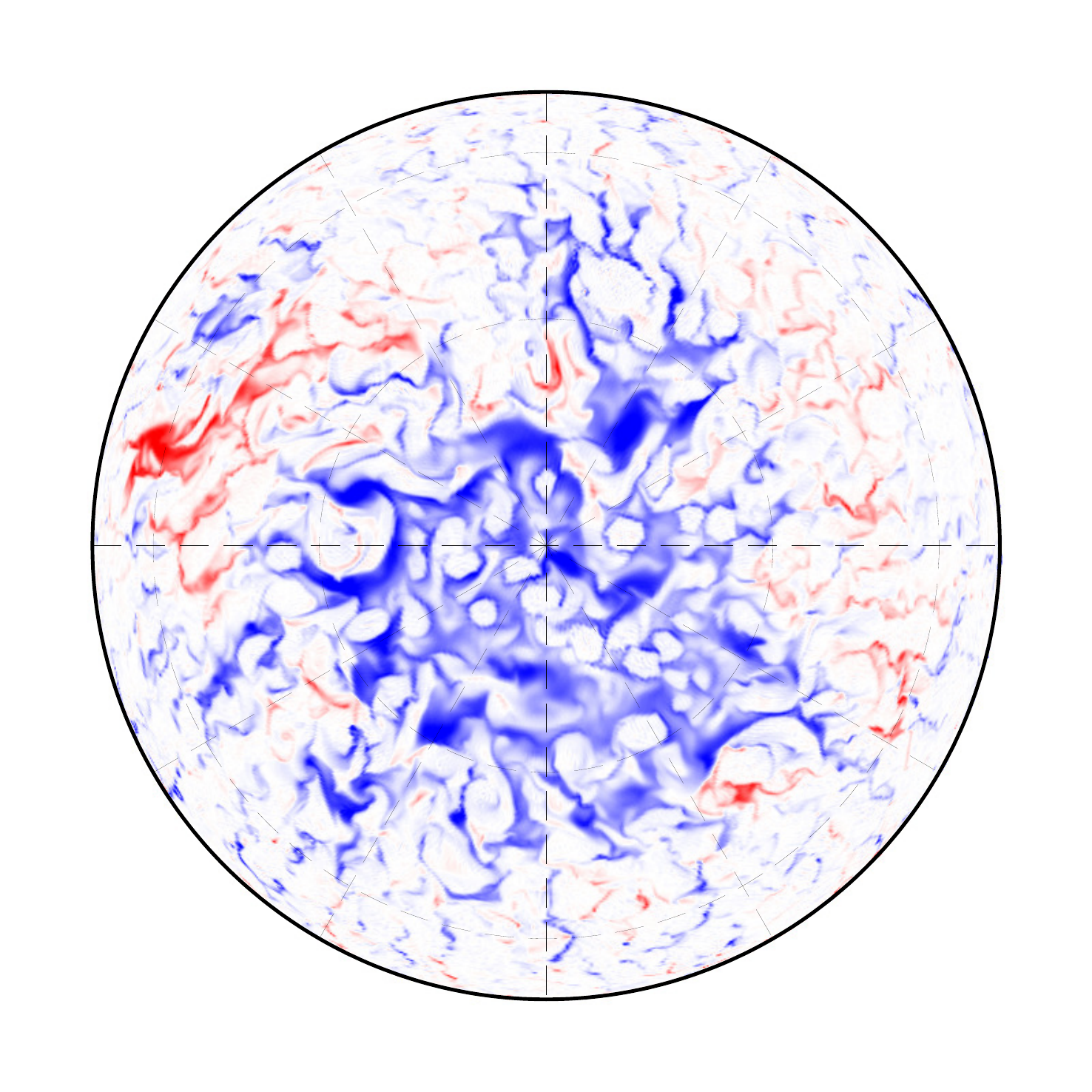}\hspace{0.5cm}
   \includegraphics[width=0.3\hsize]{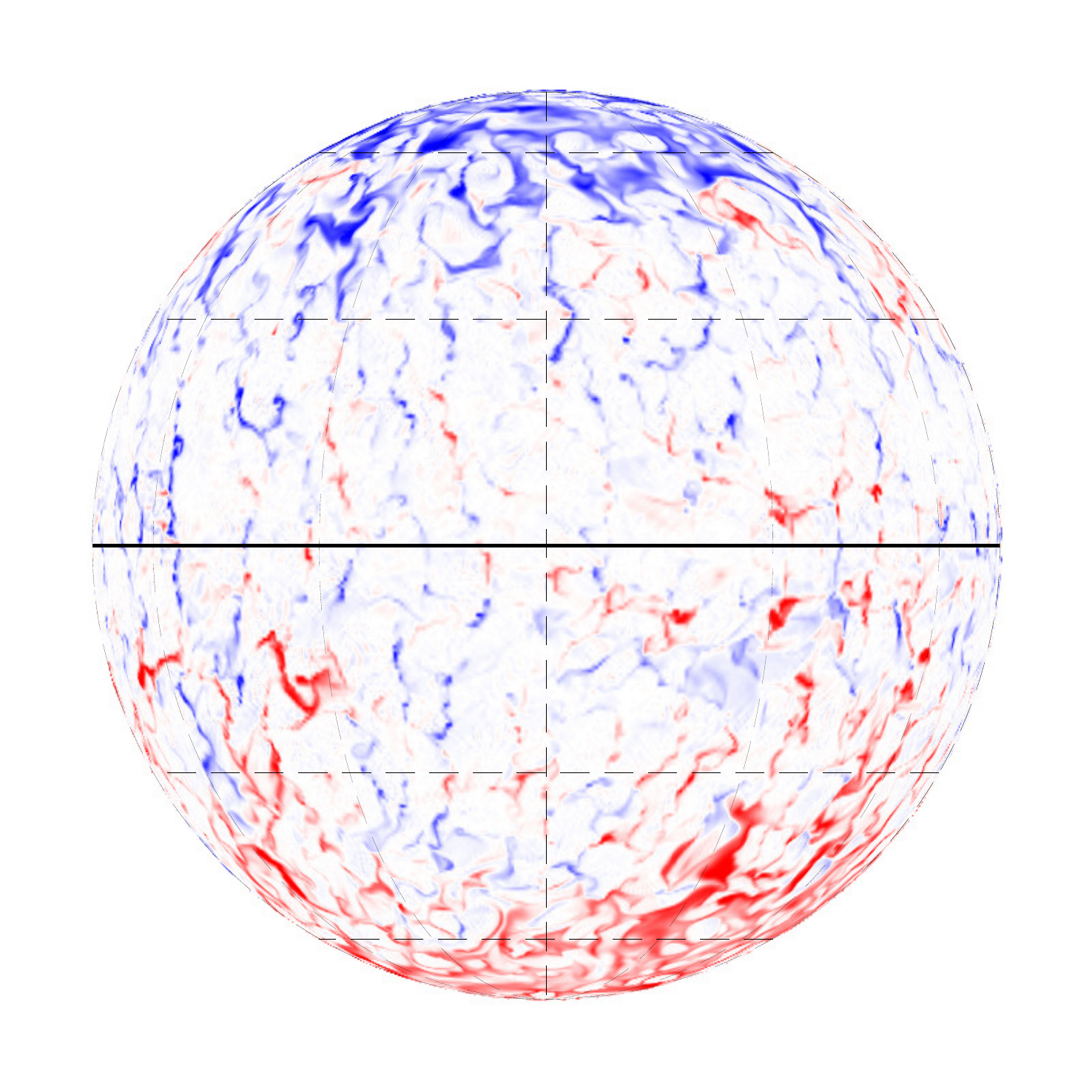}
  }
  \suppfigcaption{
   \label{fig:3d-mhd}
    \textbf{Distribution of magnetic fields from a 3D MHD model.} 
    \small
    This figure shows surface maps of radial magnetic field
    component resulted from a dedicated MHD model\supercite{2015ApJ...813L..31Y}. 
    Different colors represent 
    magnetic regions with different magnetic field orientation. It is seen that the large scale magnetic field
    of the star is dominated by a dipole with concentrations of magnetic field at polar regions.
    The two columns correspond
    to the inclination angles of $i=0$\degr\ (left) and $i=90$\degr\ (right), respectively.
    Due to a simple geometrical effect, an external observer would measure stronger magnetic field if the star is seen pole-on
    ($i=0$\degr\, $\bs=1.95$~kG) compared to the opposite case when the star is seen equator-on ($i=90$\degr, $\bs=1.37$~kG).
    The magnetic field density averaged over the whole surface of the star is $B_{\rm tot}=1.58$~kG.
    Note that the large difference in the mean magnetic field strength that corresponds to different inclination angles
    indicates that the large-scale topology of the dynamo model is not purely dipolar but includes a significant octupole contribution,
    i.e., with regions of strong magnetic field more concentrated close to the poles than in the pure dipole configuration.
}
\end{figure}

\begin{figure}
  \centerline{
    \includegraphics[width=0.9\hsize]{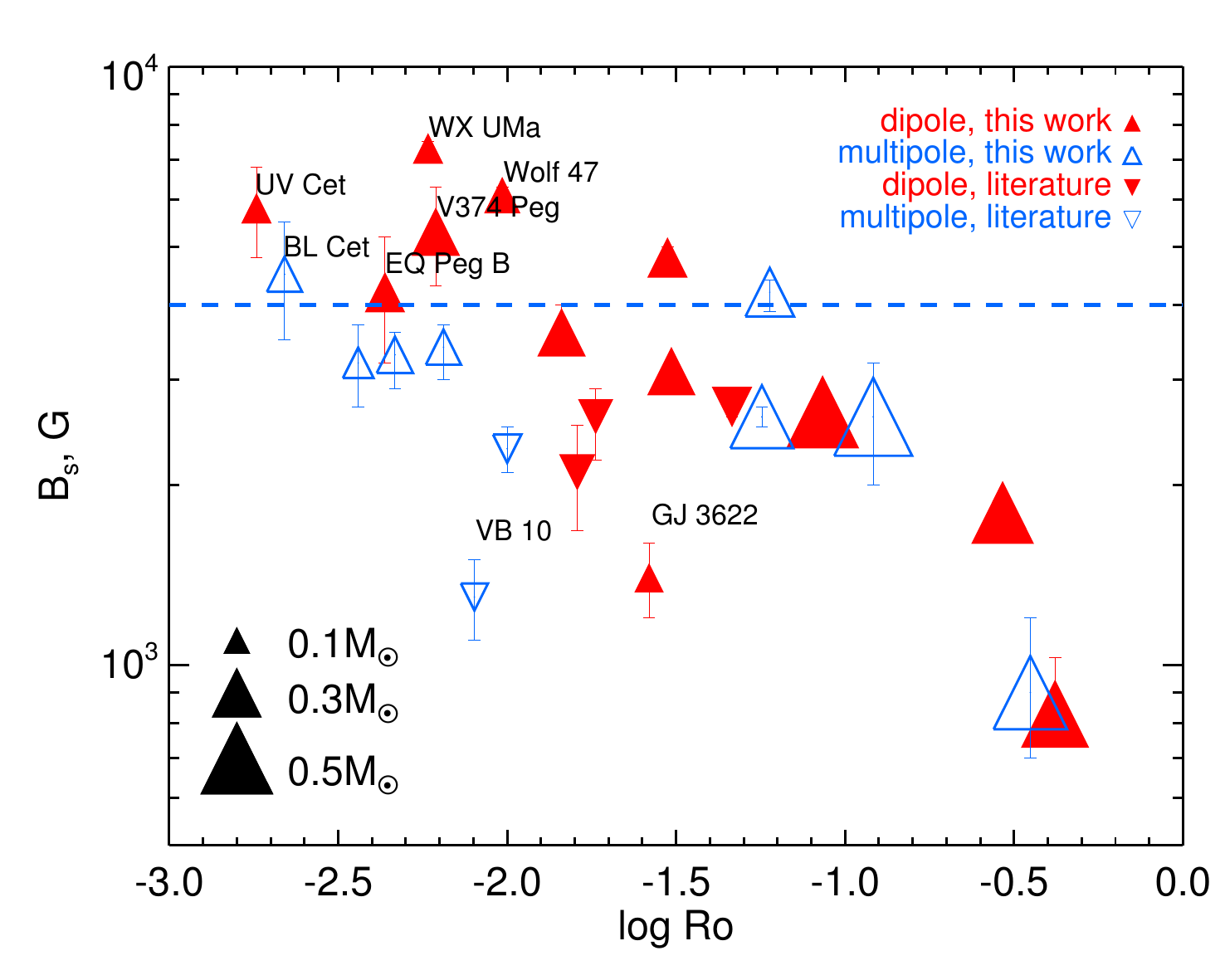}
  }
   \suppfigcaption{
   \label{fig:bf-logro}
    \textbf{Magnetic fields in M dwarfs as a function of Rossby number.} 
    \small
    We plot our measurements of stellar magentic fields as a function of Rossby number.
    The color coding and symbols are the same as in Fig.~\ref{fig:bf-period}. 
    The symbol size scales with stellar mass (see legend on the plot).
    The error bars on literature values are taken from original papers (see Supplementary  Table~\ref{tab:saturn}),
    while errors on our measurements represent estimated uncertainty rather than formal fitting errors (see Supplementary  Table~\ref{tab:magnetic}).
}
\end{figure}


\begin{figure}
  \centerline{
    \includegraphics[width=\hsize]{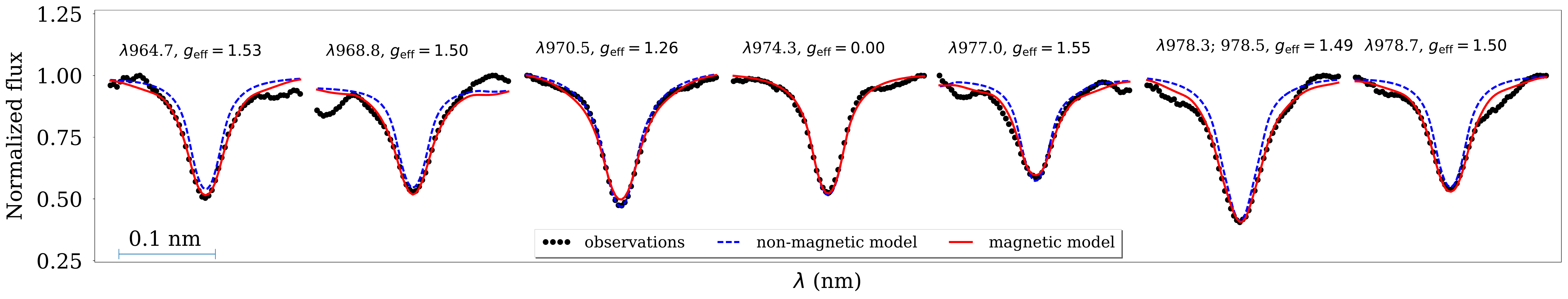}
  }
  \centerline{
    \includegraphics[width=\hsize]{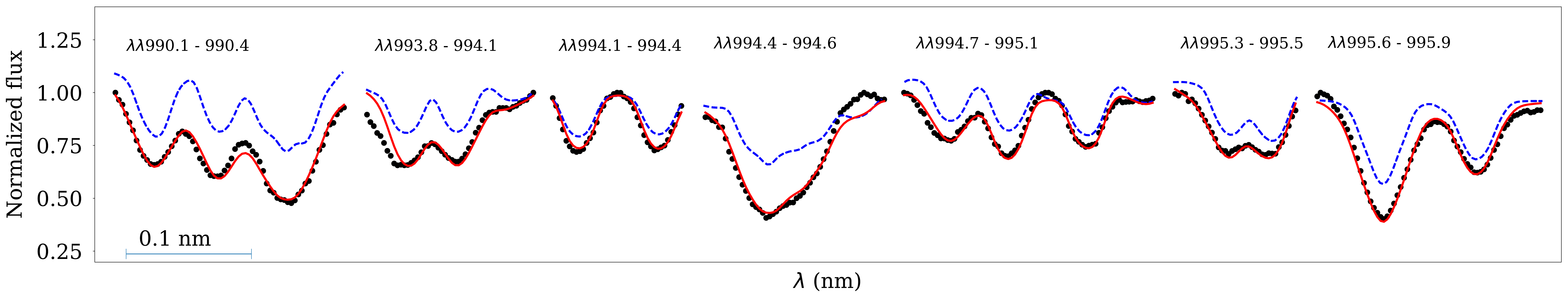}
  }
  \suppfigcaption{
   \label{fig:fit-dxcnc}
    \textbf{Model fit to Ti and FeH lines in DX~Cnc.} 
    \small
    We show the comparison between observed and predicted spectra
    for a set of Ti (top panel) and FeH lines (bottom panel) for the model
    with $\teff=2900$~K.
    Black diamonds~--~observations; red full line~--~best fit model spectrum;
    blue dashed line~--~spectrum computed assuming zero magnetic field.
    Corresponding atmospheric parameters and magnetic field are listed in
    Supplementary  Table~\ref{tab:magnetic}.
}
\end{figure}

\begin{figure}
  \centerline{
    \includegraphics[width=\hsize]{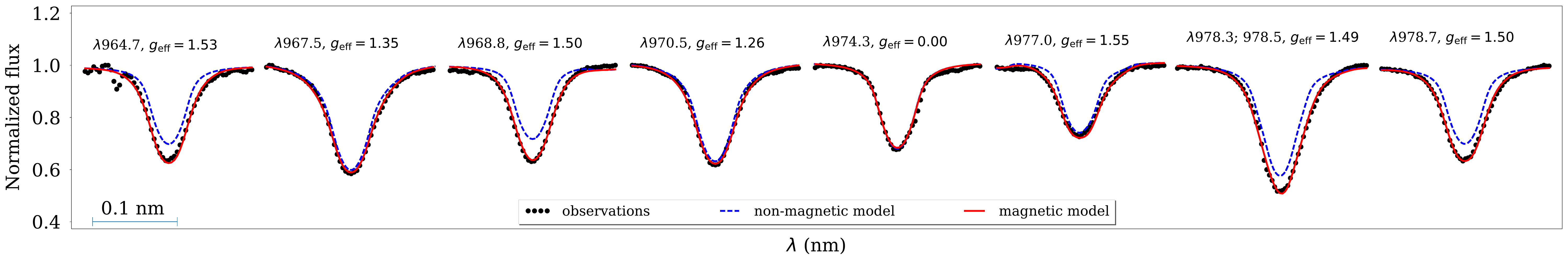}
  }
  \centerline{
    \includegraphics[width=\hsize]{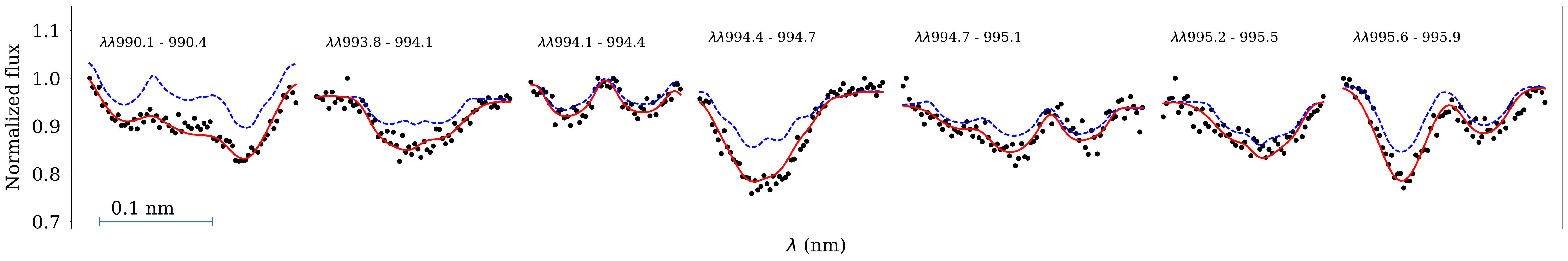}
  }
  \suppfigcaption{
   \label{fig:fit-eqpega}
    \textbf{Same as in Supplementary  Fig.~\ref{fig:fit-dxcnc} but for EQ~Peg~A.} 
    \small
    We show fit for $\teff=3400$~K model.
}
\end{figure}

\clearpage
\begin{figure}
  \centerline{
    \includegraphics[width=\hsize]{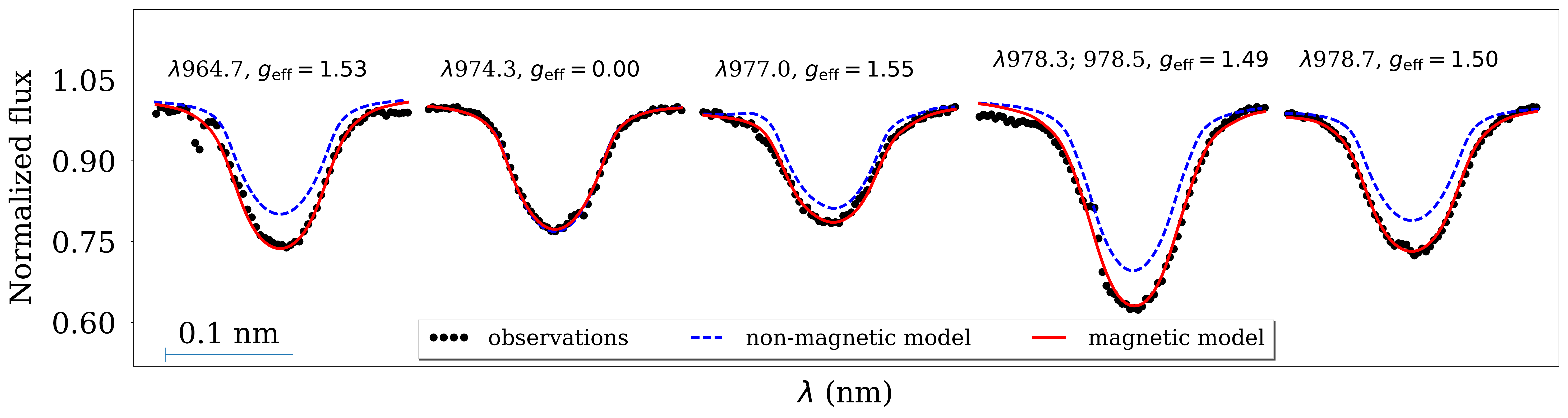}
  }
  \centerline{
    \includegraphics[width=\hsize]{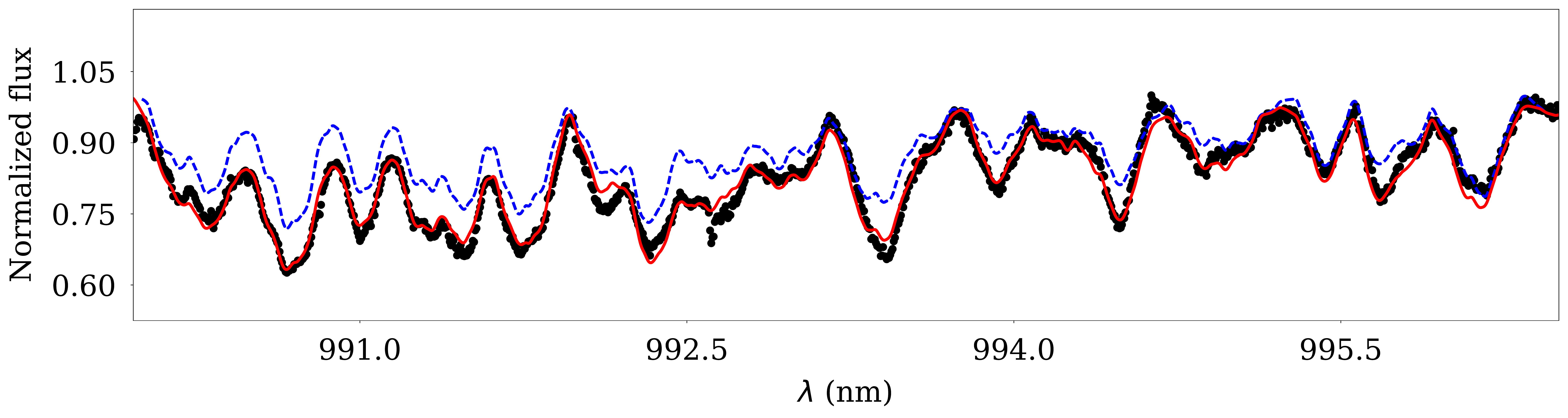}
  }
  \suppfigcaption{
   \label{fig:fit-eqpegb}
    \textbf{Same as in Supplementary  Fig.~\ref{fig:fit-dxcnc} but for EQ~Peg~B.} 
    \small
    We show fit for $\teff=3300$~K model.
}
\end{figure}

\begin{figure}
  \centerline{
    \includegraphics[width=\hsize]{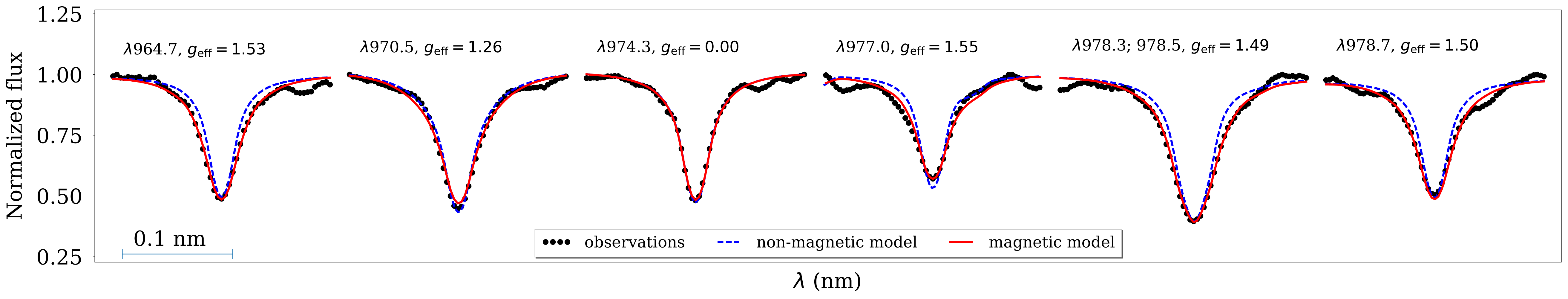}
  }
  \centerline{
    \includegraphics[width=\hsize]{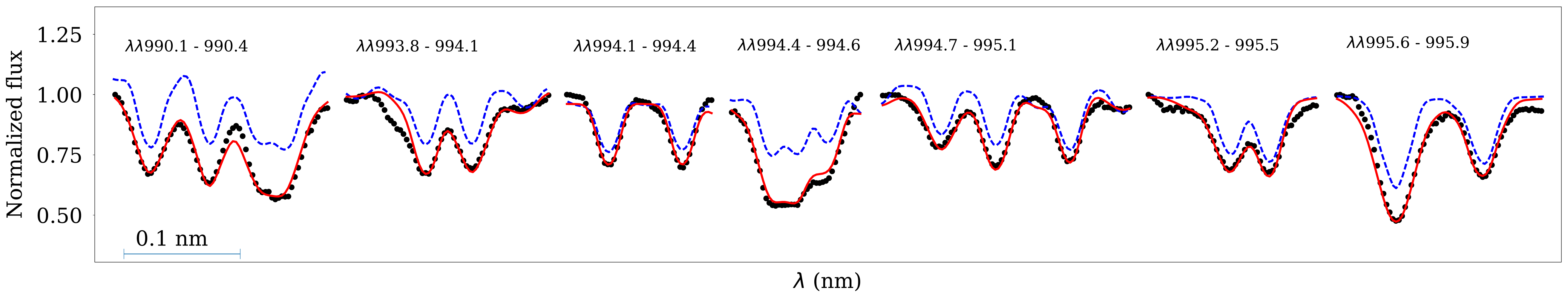}
  }
  \suppfigcaption{
   \label{fig:fit-gj1245b}
    \textbf{Same as in Supplementary  Fig.~\ref{fig:fit-dxcnc} but for GJ~1245~B.} 
    \small
    We show fit for $\teff=3100$~K model.
}
\end{figure}

\clearpage
\begin{figure}
  \centerline{
    \includegraphics[width=\hsize]{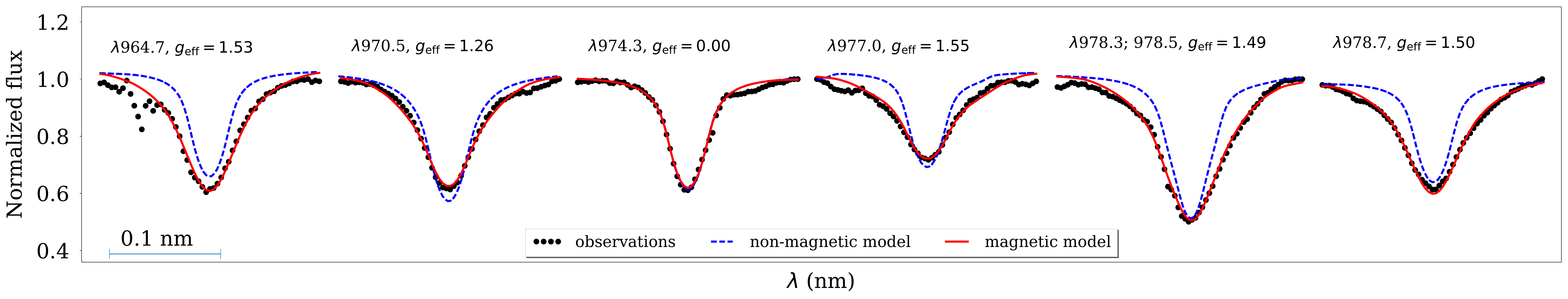}
  }
  \centerline{
    \includegraphics[width=\hsize]{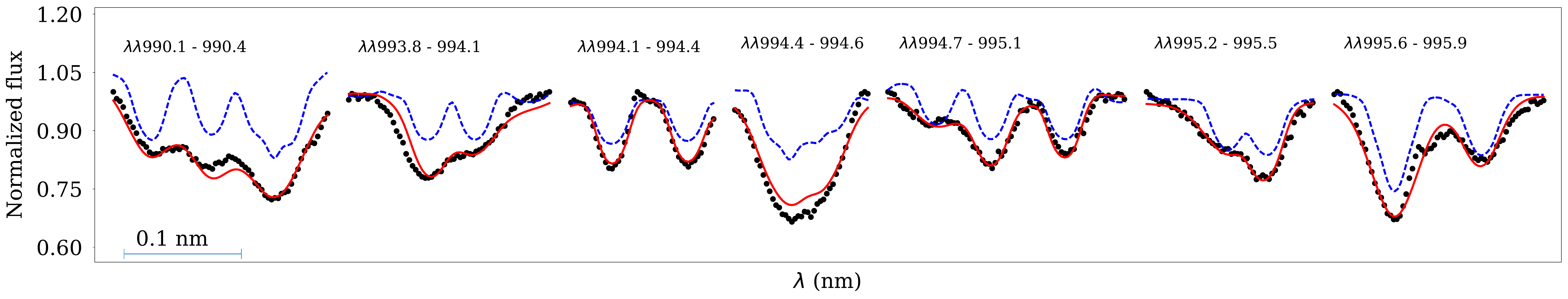}
  }
  \suppfigcaption{
   \label{fig:fit-gj51}
    \textbf{Same as in Supplementary  Fig.~\ref{fig:fit-dxcnc} but for Gl~51.} 
    \small
    We show fit for $\teff=3200$~K model.
}
\end{figure}

\begin{figure}
  \centerline{
    \includegraphics[width=\hsize]{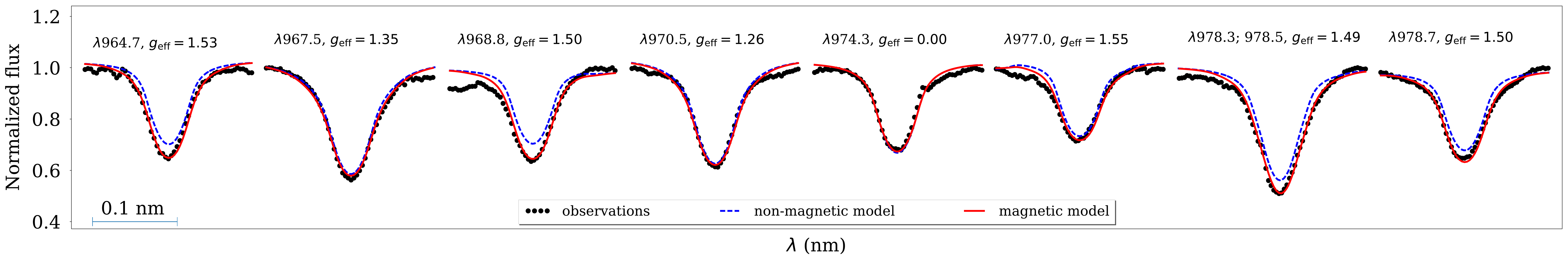}
  }
  \centerline{
    \includegraphics[width=\hsize]{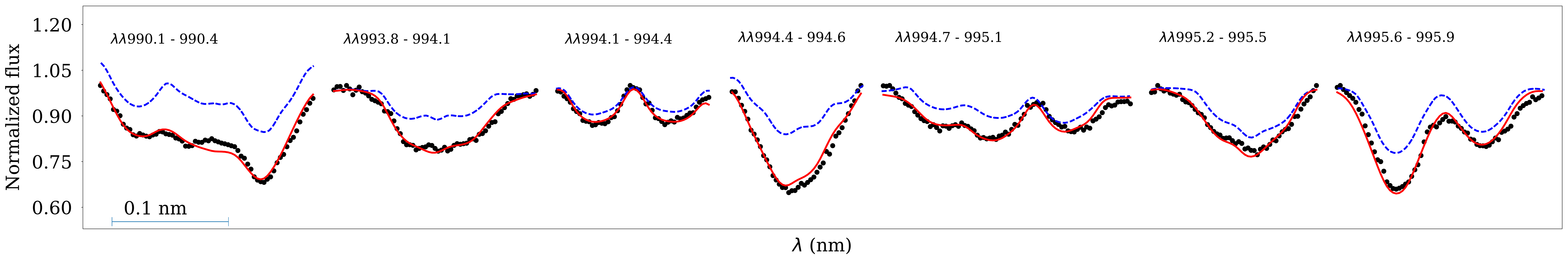}
  }
  \suppfigcaption{
   \label{fig:fit-glvir}
    \textbf{Same as in Supplementary  Fig.~\ref{fig:fit-dxcnc} but for Gl~Vir.} 
    \small
    We show fit for $\teff=3200$~K model.
}
\end{figure}

\clearpage
\begin{figure}
  \centerline{
    \includegraphics[width=\hsize]{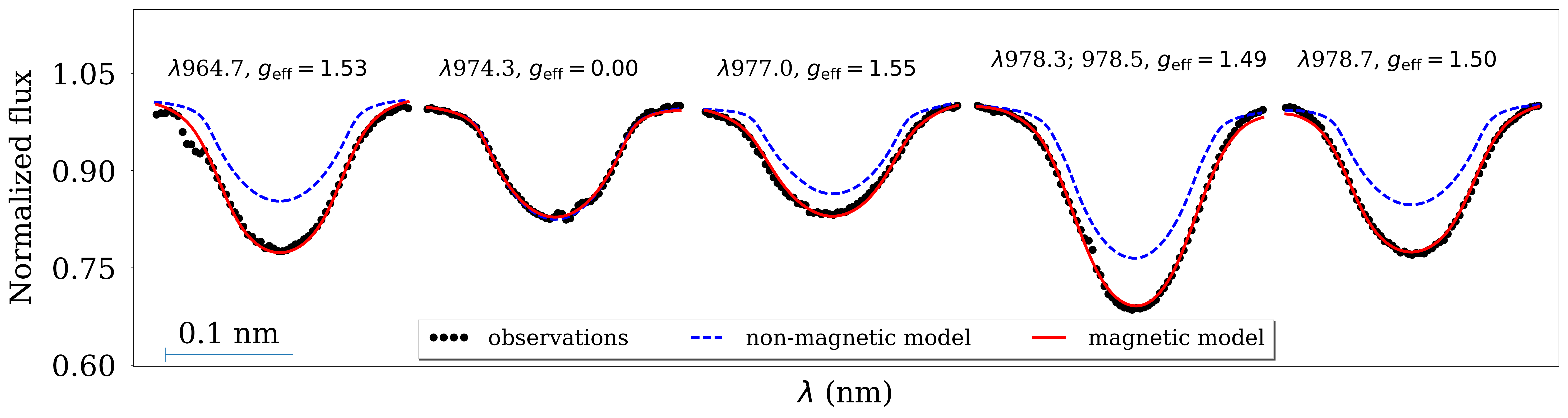}
  }
  \centerline{
    \includegraphics[width=\hsize]{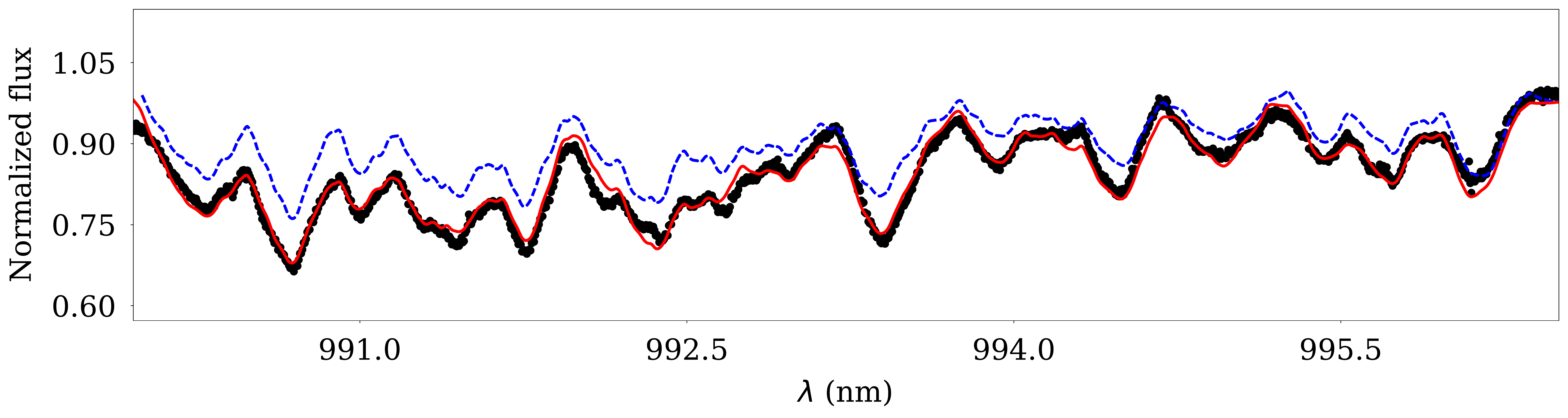}
  }
  \suppfigcaption{
   \label{fig:fit-v374peg}
    \textbf{Same as in Supplementary  Fig.~\ref{fig:fit-dxcnc} but for V374~Peg.} 
    \small
    We show fit for $\teff=3400$~K model.
}
\end{figure}

\begin{figure}
  \centerline{
    \includegraphics[width=\hsize]{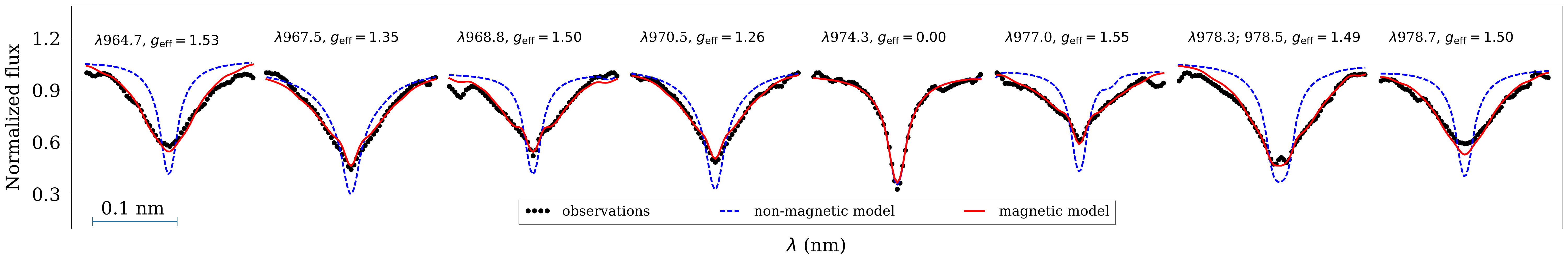}
  }
  \centerline{
    \includegraphics[width=\hsize]{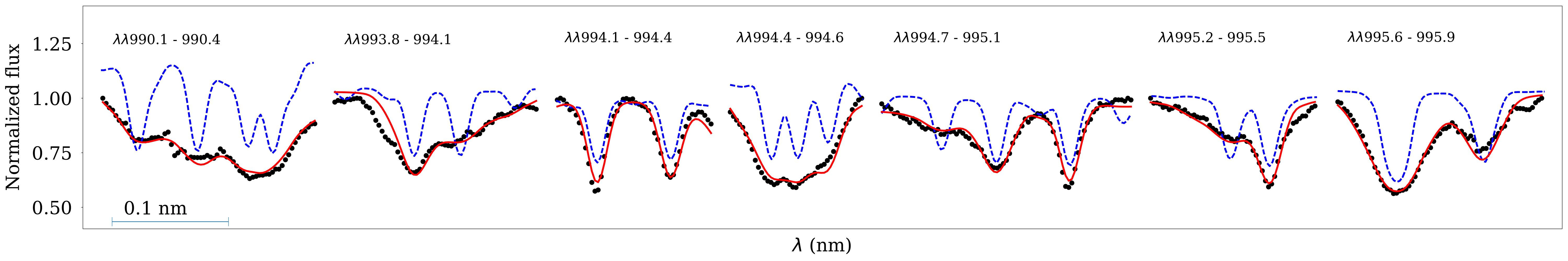}
  }
  \suppfigcaption{
   \label{fig:fit-wxuma}
    \textbf{Same as in Supplementary  Fig.~\ref{fig:fit-dxcnc} but for WX~UMa.} 
    \small
    We show fit for $\teff=2900$~K model.
}
\end{figure}

\clearpage
\begin{figure}
  \centerline{
    \includegraphics[width=\hsize]{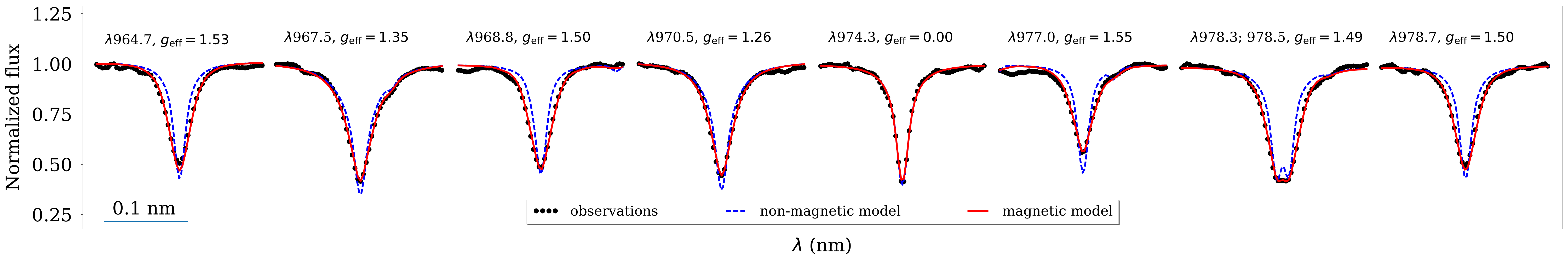}
  }
  \centerline{
    \includegraphics[width=\hsize]{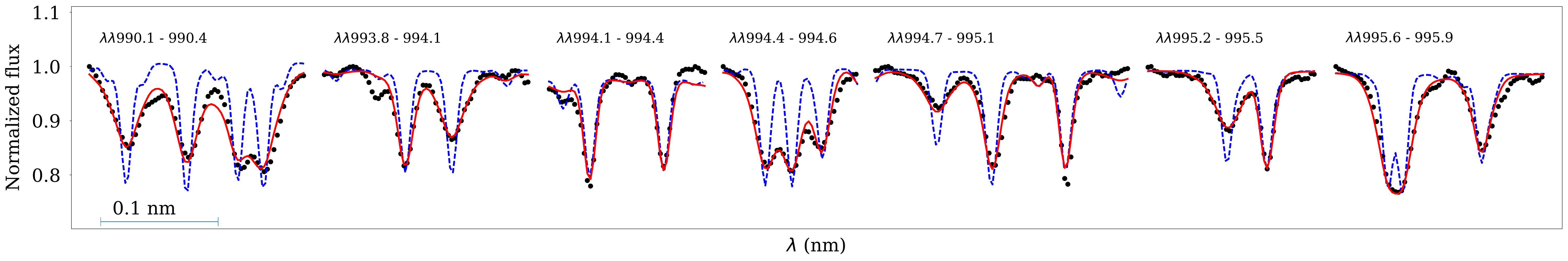}
  }
  \suppfigcaption{
   \label{fig:fit-adleo}
    \textbf{Same as in Supplementary  Fig.~\ref{fig:fit-dxcnc} but for AD~Leo.} 
    \small
    We show fit for $\teff=3300$~K model.
}
\end{figure}

\begin{figure}
  \centerline{
    \includegraphics[width=\hsize]{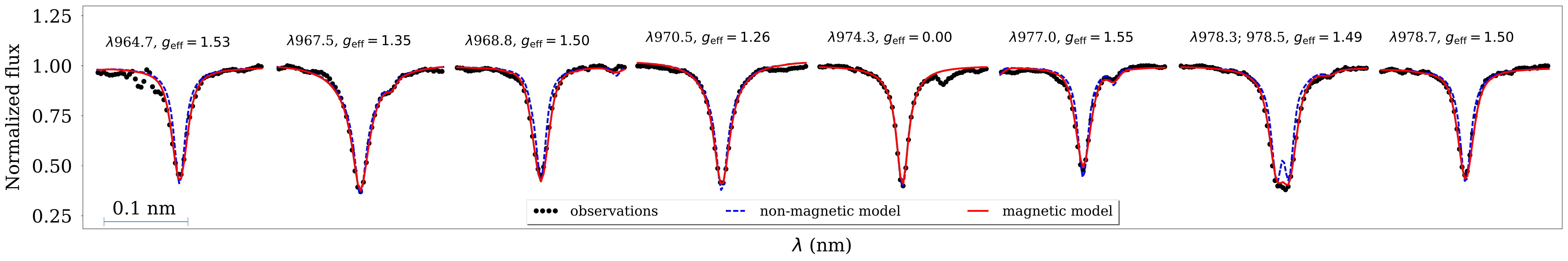}
  }
  \centerline{
    \includegraphics[width=\hsize]{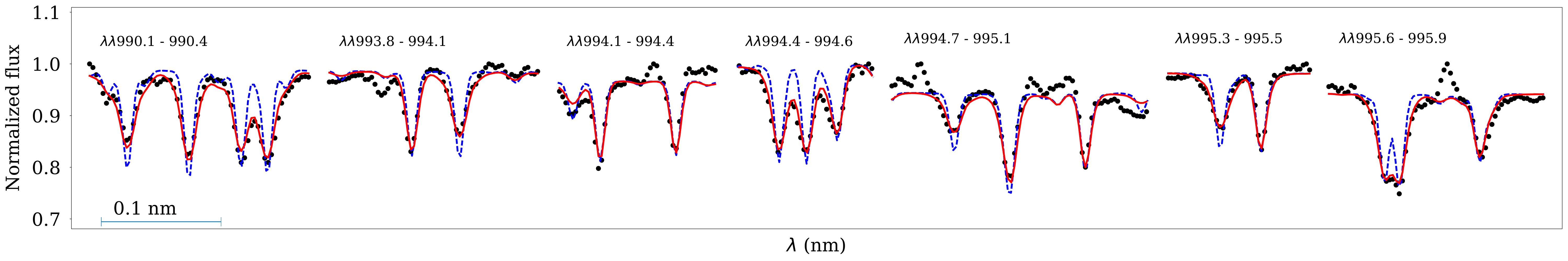}
  }
  \suppfigcaption{
   \label{fig:fit-ceboo}
    \textbf{Same as in Supplementary  Fig.~\ref{fig:fit-dxcnc} but for CE~Boo.} 
    \small
    We show fit for $\teff=3500$~K model.
}
\end{figure}

\clearpage
\begin{figure}
  \centerline{
    \includegraphics[width=\hsize]{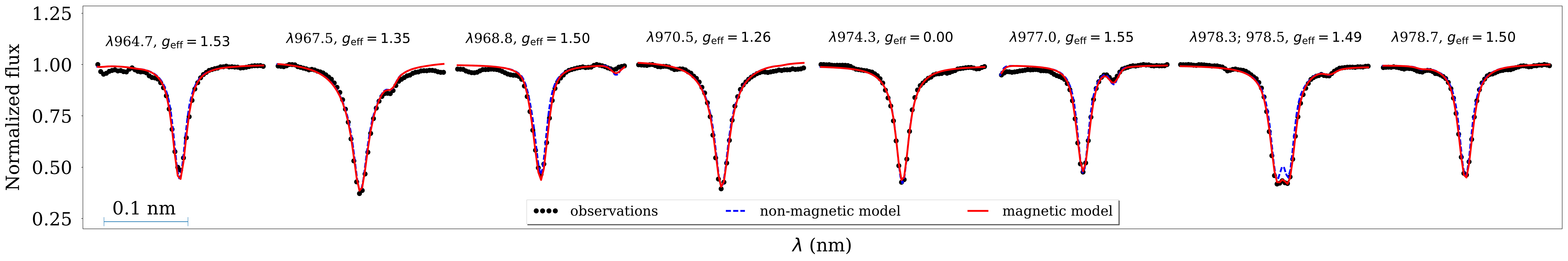}
  }
  \centerline{
    \includegraphics[width=\hsize]{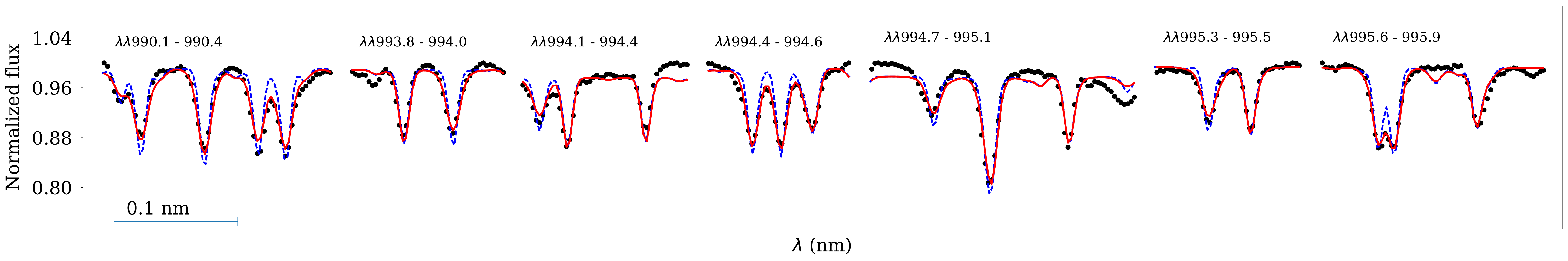}
  }
  \suppfigcaption{
   \label{fig:fit-dsleo}
    \textbf{Same as in Supplementary  Fig.~\ref{fig:fit-dxcnc} but for DS~Leo.} 
    \small
    We show fit for $\teff=3600$~K model.
}
\end{figure}

\begin{figure}
  \centerline{
    \includegraphics[width=\hsize]{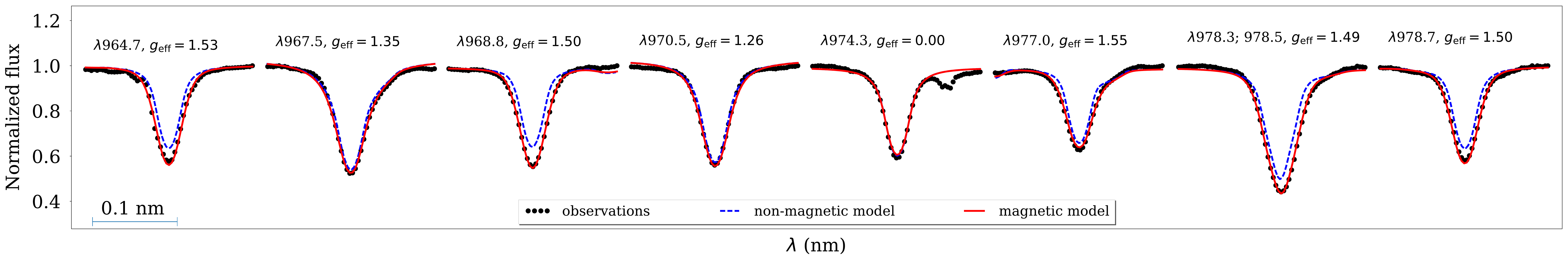}
  }
  \centerline{
    \includegraphics[width=\hsize]{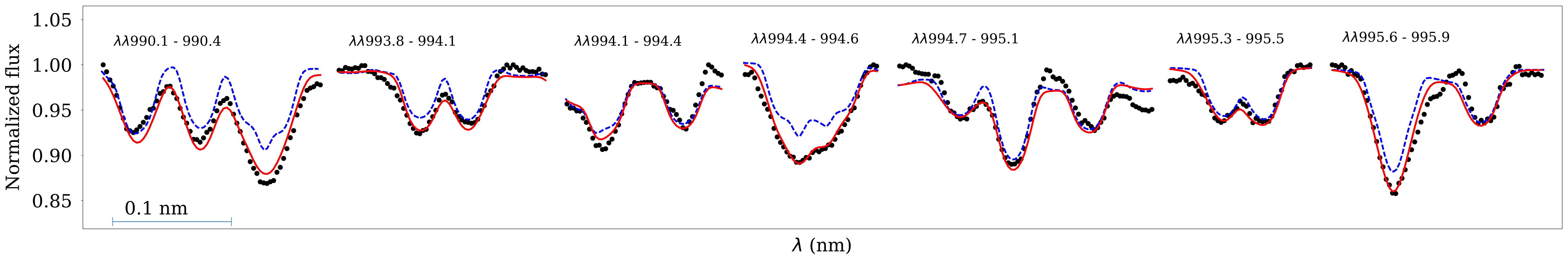}
  }
  \suppfigcaption{
   \label{fig:fit-dtvir}
    \textbf{Same as in Supplementary  Fig.~\ref{fig:fit-dxcnc} but for DT~Vir.} 
    \small
    We show fit for $\teff=3600$~K model.
}
\end{figure}

\clearpage
\begin{figure}
  \centerline{
    \includegraphics[width=\hsize]{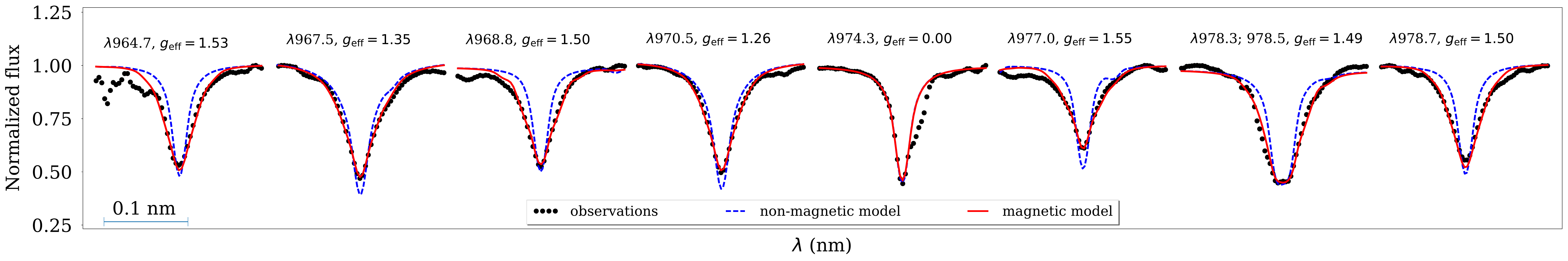}
  }
  \centerline{
    \includegraphics[width=\hsize]{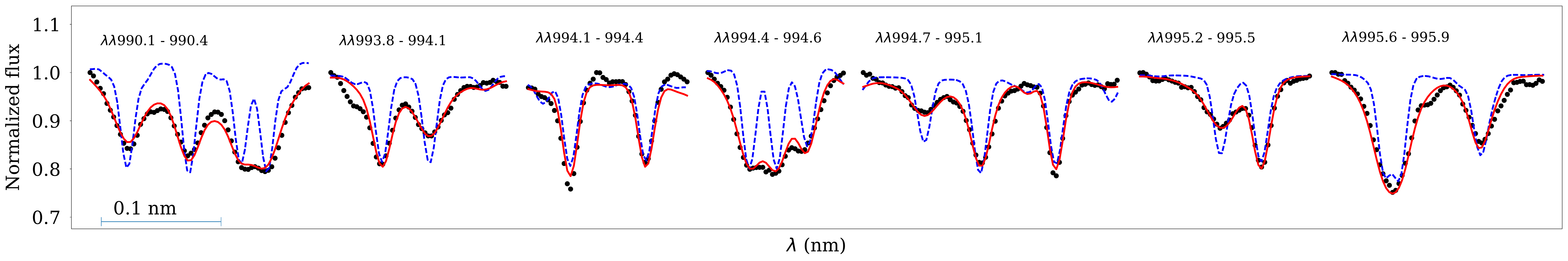}
  }
  \suppfigcaption{
   \label{fig:fit-evlac}
    \textbf{Same as in Supplementary  Fig.~\ref{fig:fit-dxcnc} but for EV~Lac.} 
    \small
    We show fit for $\teff=3200$~K model.
}
\end{figure}

\begin{figure}
  \centerline{
    \includegraphics[width=\hsize]{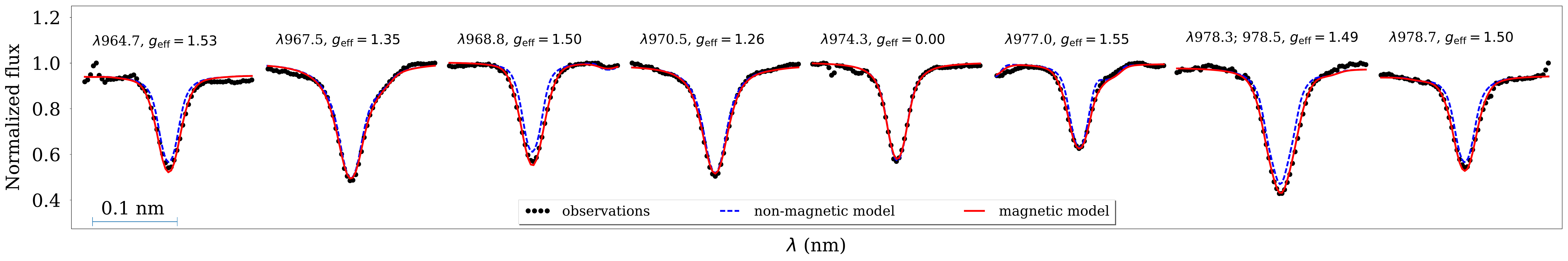}
  }
  \centerline{
    \includegraphics[width=\hsize]{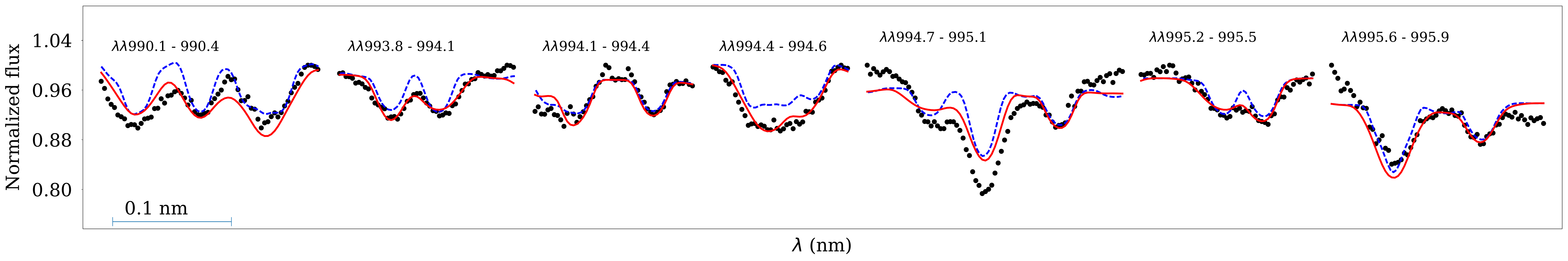}
  }
  \suppfigcaption{
   \label{fig:fit-gj182}
    \textbf{Same as in Supplementary  Fig.~\ref{fig:fit-dxcnc} but for Gl~182.} 
    \small
    We show fit for $\teff=3800$~K model.
}
\end{figure}

\clearpage
\begin{figure}
  \centerline{
    \includegraphics[width=\hsize]{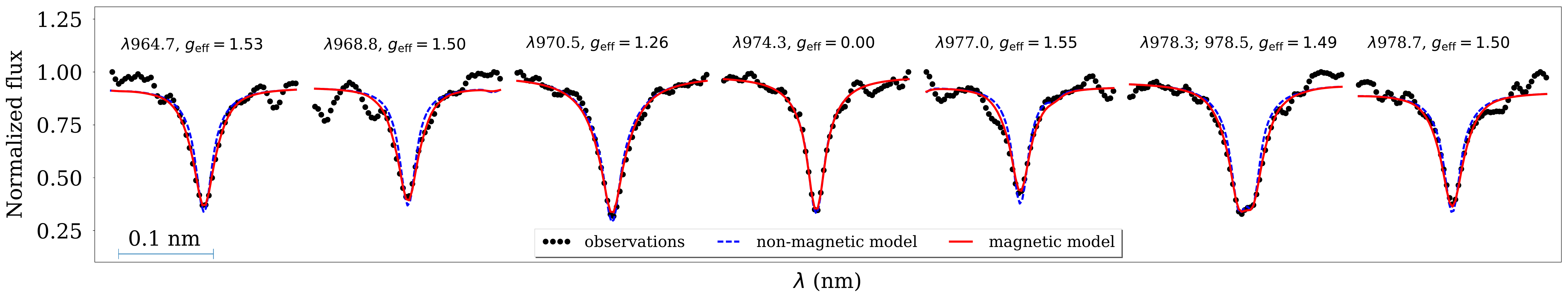}
  }
  \centerline{
    \includegraphics[width=\hsize]{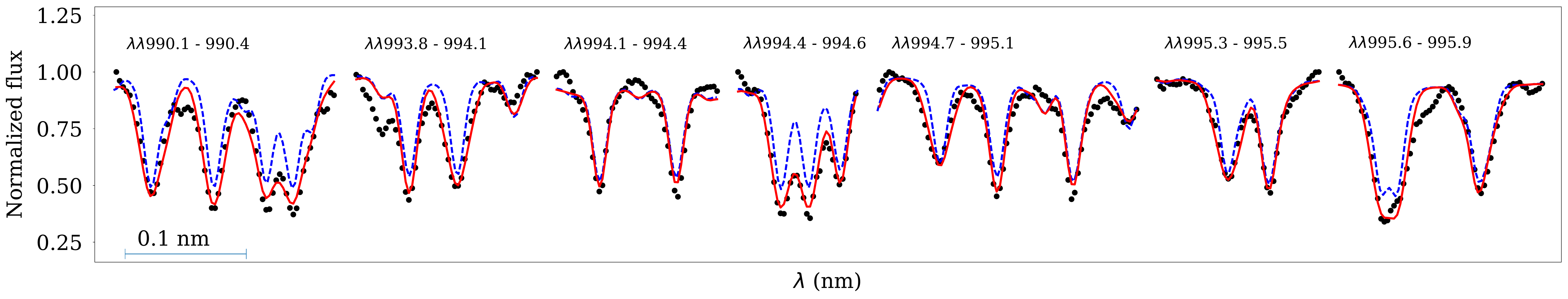}
  }
  \suppfigcaption{
   \label{fig:fit-gj3622}
    \textbf{Same as in Supplementary  Fig.~\ref{fig:fit-dxcnc} but for GJ~3622.} 
    \small
    We show fit for $\teff=2800$~K model.
}
\end{figure}

\begin{figure}
  \centerline{
    \includegraphics[width=\hsize]{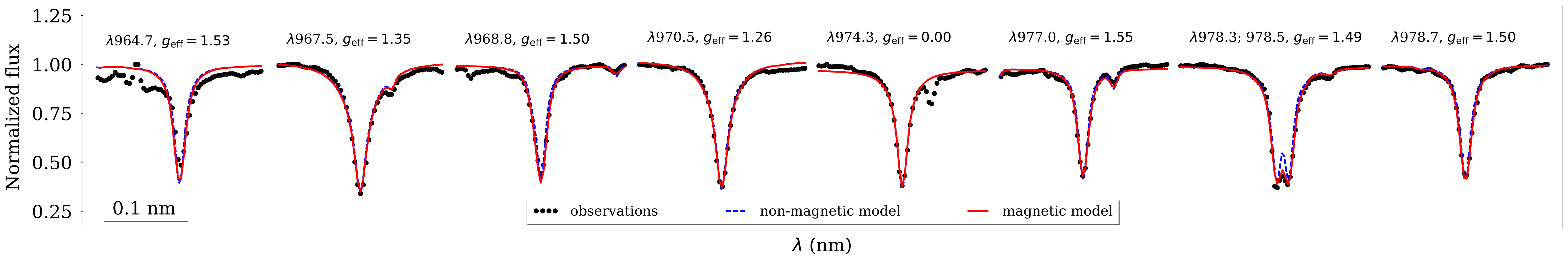}
  }
  \centerline{
    \includegraphics[width=\hsize]{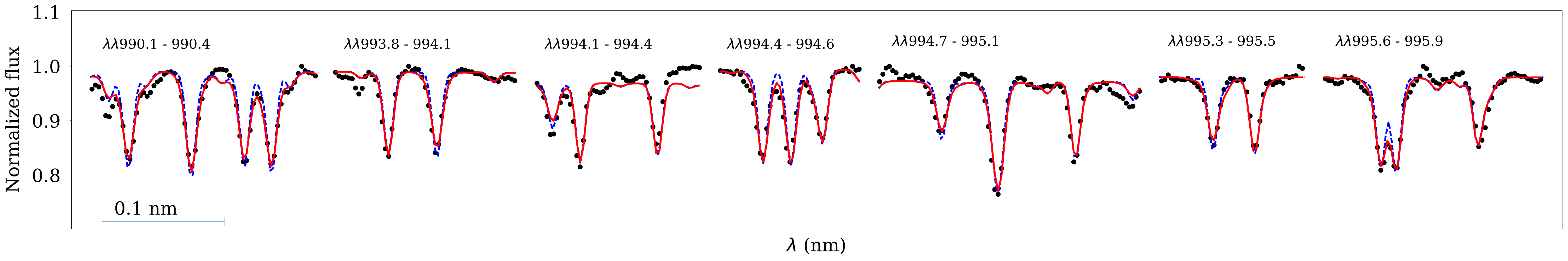}
  }
  \suppfigcaption{
   \label{fig:fit-gj49}
    \textbf{Same as in Supplementary  Fig.~\ref{fig:fit-dxcnc} but for Gl~49.} 
    \small
    We show fit for $\teff=3600$~K model.
}
\end{figure}

\clearpage
\begin{figure}
  \centerline{
    \includegraphics[width=\hsize]{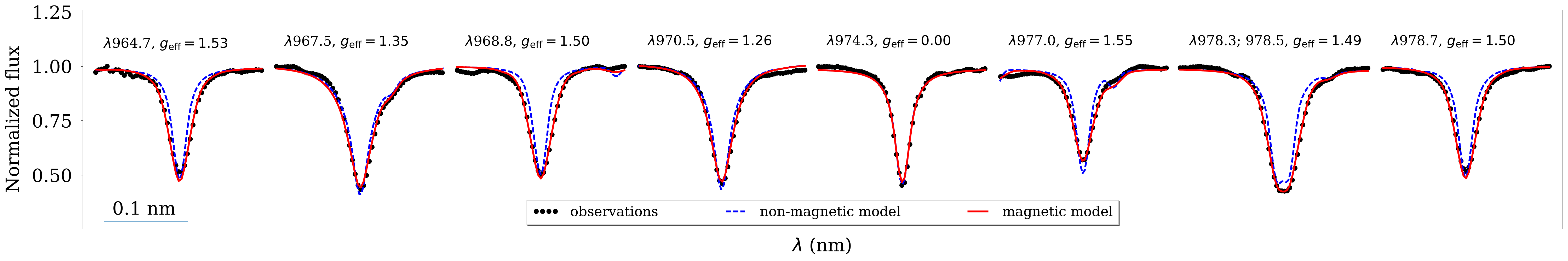}
  }
  \centerline{
    \includegraphics[width=\hsize]{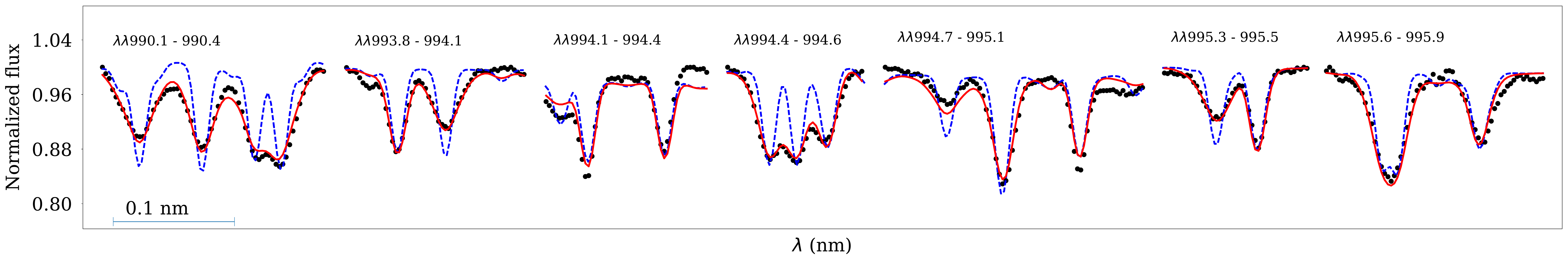}
  }
  \suppfigcaption{
   \label{fig:fit-otser}
    \textbf{Same as in Supplementary  Fig.~\ref{fig:fit-dxcnc} but for OT~Ser.} 
    \small
    We show fit for $\teff=3600$~K model.
}
\end{figure}

\begin{figure}
  \centerline{
    \includegraphics[width=\hsize]{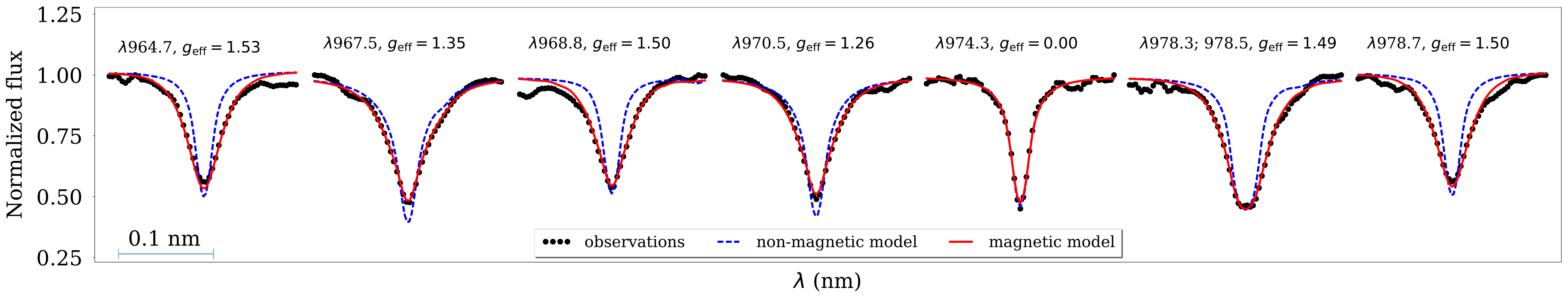}
  }
  \centerline{
    \includegraphics[width=\hsize]{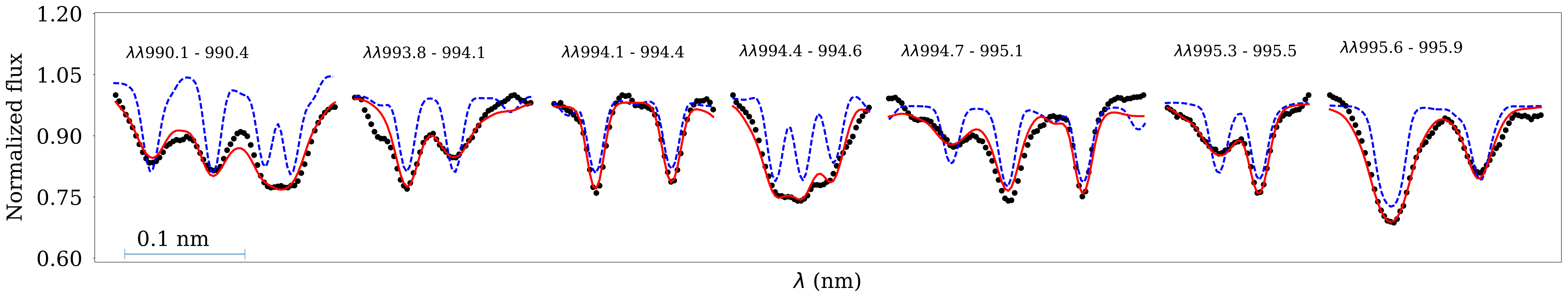}
  }
  \suppfigcaption{
   \label{fig:fit-yzcmi}
    \textbf{Same as in Supplementary  Fig.~\ref{fig:fit-dxcnc} but for YZ~CMi.} 
    \small
    We show fit for $\teff=3200$~K model.
}
\end{figure}

\clearpage
\begin{figure}
  \centerline{
    \includegraphics[width=\hsize]{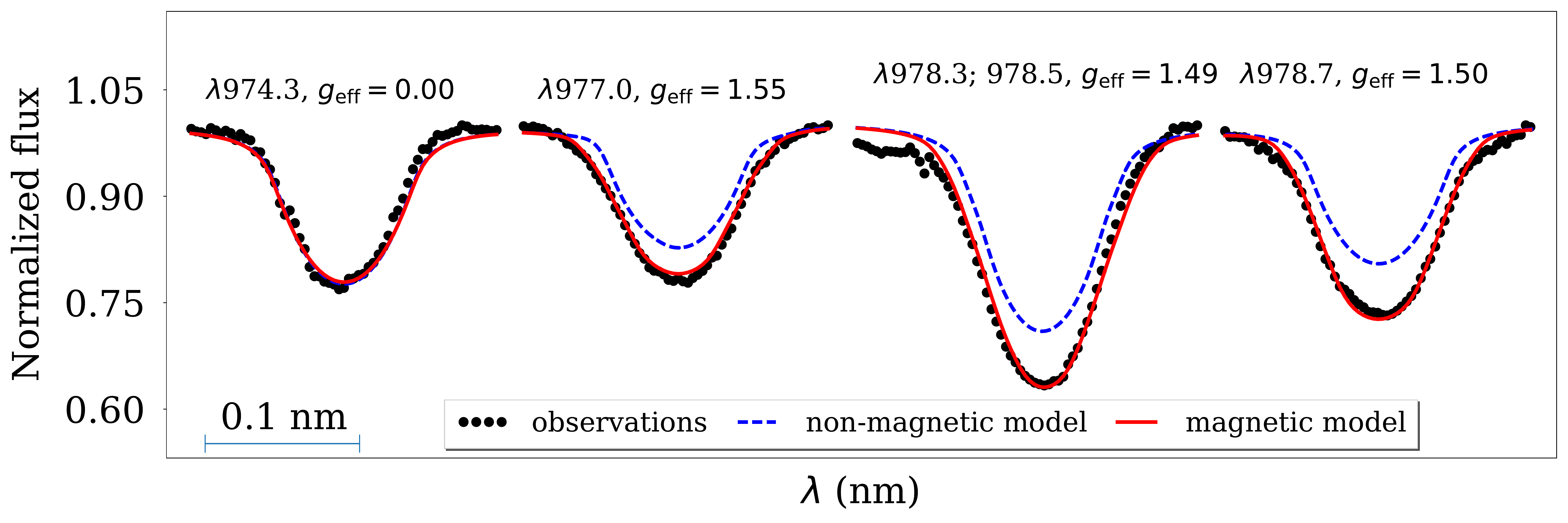}
  }
  \centerline{
    \includegraphics[width=\hsize]{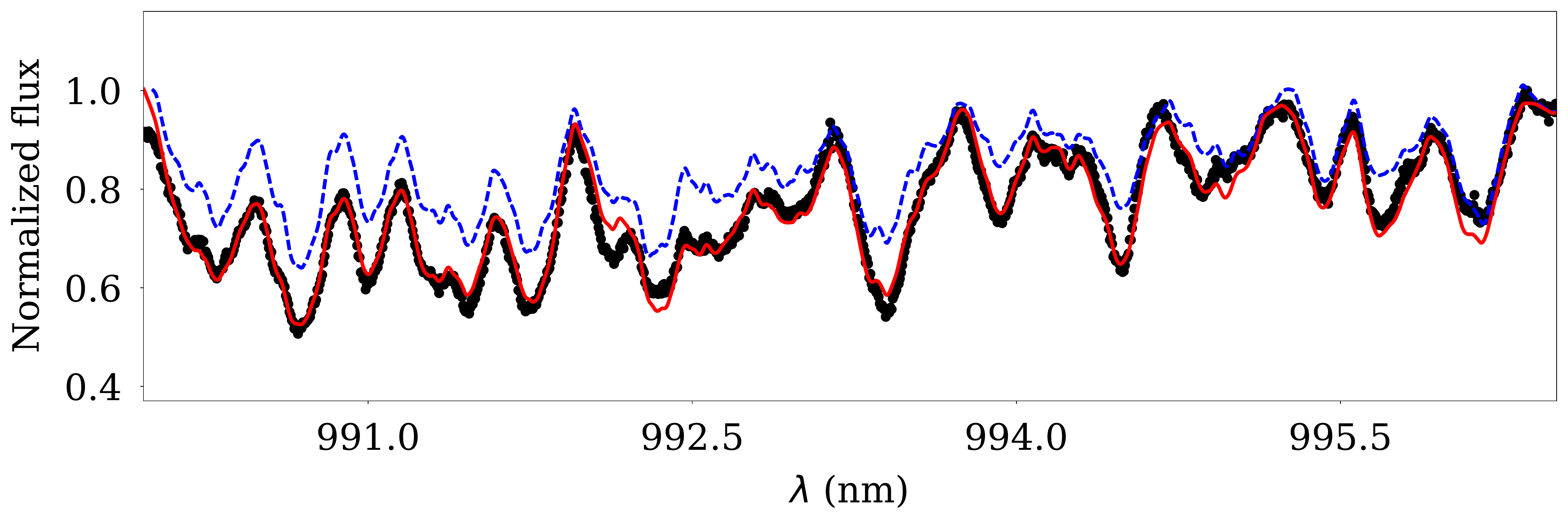}
  }
  \suppfigcaption{
   \label{fig:fit-gj65a}
    \textbf{Same as in Supplementary  Fig.~\ref{fig:fit-dxcnc} but for Gl~65~A.} 
    \small
    We show fit for $\teff=3000$~K model.
}
\end{figure}

\begin{figure}
  \centerline{
    \includegraphics[width=\hsize]{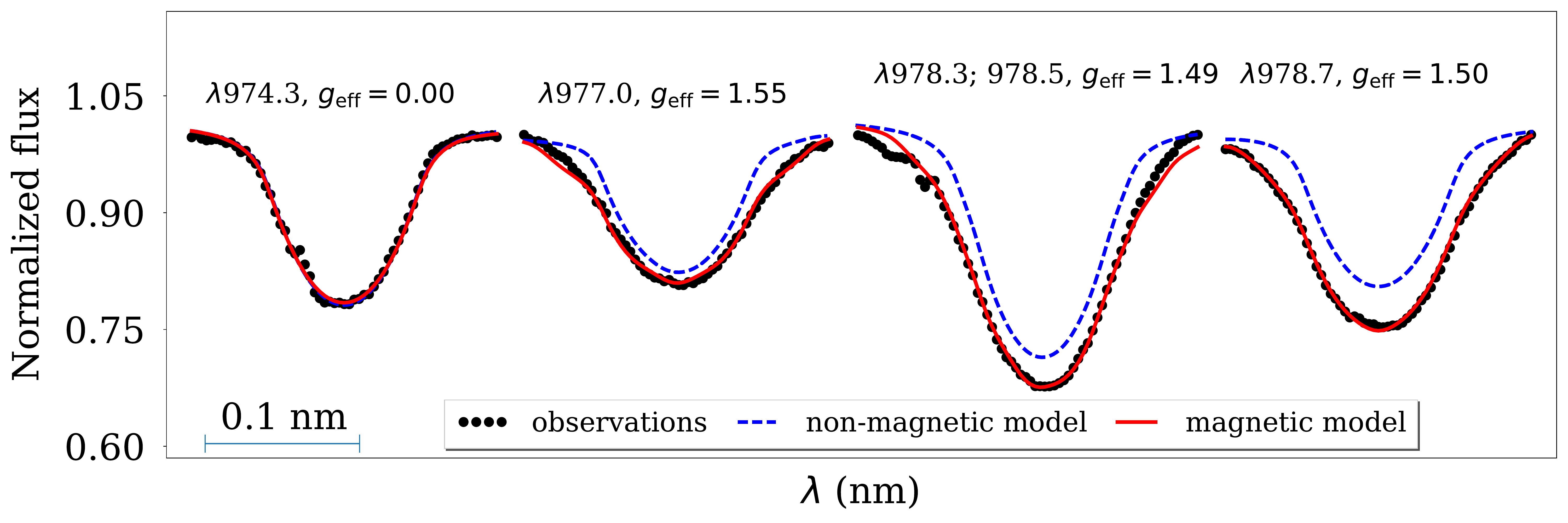}
  }
  \centerline{
    \includegraphics[width=\hsize]{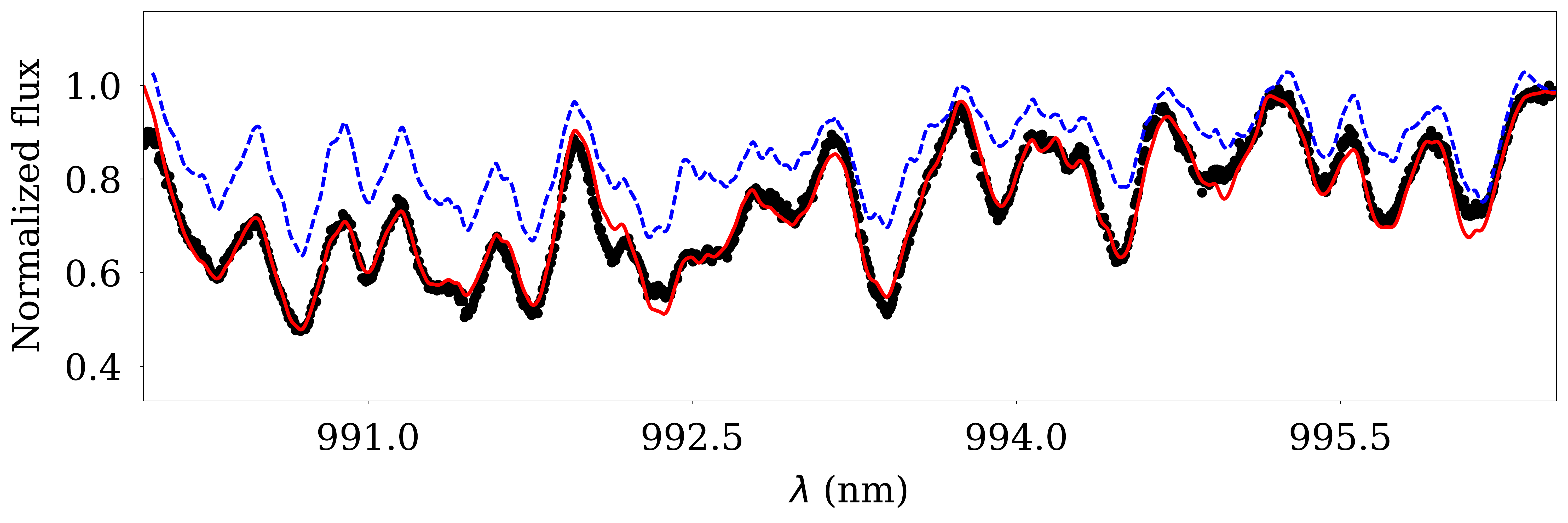}
  }
  \suppfigcaption{
   \label{fig:fit-gj65b}
    \textbf{Same as in Supplementary  Fig.~\ref{fig:fit-dxcnc} but for Gl~65~B.} 
    \small
    We show fit for $\teff=2900$~K model.
}
\end{figure}

\clearpage

\printbibliography[segment=\therefsegment,resetnumbers=false,filter=notmethods,filter=notmain]
\end{refsegment}

\end{supplementaryinformation}

\end{document}